\documentclass[showpacs,prd,nofootinbib,reprint,preprintnumbers]{revtex4-1}
\usepackage{adjustbox}
\usepackage[utf8]{inputenc}
\usepackage{color,graphicx,epsfig}
\usepackage{amsmath,amssymb,dsfont,epsfig,graphicx,xcolor,fontenc}
\usepackage{bm}
\usepackage[english]{babel}
\usepackage{graphicx}%
\usepackage{amsfonts}%
\usepackage{amssymb}
\usepackage{braket}
\usepackage{hyperref}
\usepackage{epstopdf}
\usepackage{ulem}
\usepackage{soul}

\definecolor{nicered}{rgb}{0.7,0.1,0.1}
\definecolor{nicegreen}{rgb}{0.1,0.5,0.1}
\definecolor{violet}{rgb}{0.7,0.3,0.3}
\hypersetup{colorlinks,citecolor= nicegreen,linkcolor= nicered}

\setstcolor{red}

\newcommand{\mc}{\mathcal}

\emergencystretch=\maxdimen
\def\LjubljanaFMF{Faculty of Mathematics and Physics, University of Ljubljana,
 Jadranska 19, 1000 Ljubljana, Slovenia }
\def\LjubljanaIJS{Jo\v zef Stefan Institute, Jamova 39, 1000 Ljubljana, Slovenia}
\def\IFIC{Instituto de F\'isica Corpuscular,
Universitat de Val\'encia – Consejo Superior de Investigaciones Cient\'ificas,
Parc Cient\'ific, E-46980 Paterna, Valencia, Spain}
\def\Bari{Istituto Nazionale di Fisica Nucleare, Sezione di Bari, Via Orabona 4, 70126 Bari, Italy}
\begin{document}
\preprint{BARI-TH/757-24}

\title{Signatures of Light New Particles in $B\to K^{(*)} E_{\rm miss}$}

\author{Patrick D. Bolton}
\email[Electronic address: ]{patrick.bolton@ijs.si} 
\affiliation{\LjubljanaIJS}

\author{Svjetlana Fajfer}
\email[Electronic address: ]{svjetlana.fajfer@ijs.si} 
\affiliation{\LjubljanaIJS}
\affiliation{\LjubljanaFMF}

\author{Jernej~F.~Kamenik}
\email[Electronic address: ]{jernej.kamenik@cern.ch} 
\affiliation{\LjubljanaIJS}
\affiliation{\LjubljanaFMF}

\author{Mart\'in Novoa-Brunet}
\email[Electronic address: ]{martin.novoa@ific.uv.es}
\affiliation{\IFIC}
\affiliation{\Bari}

\begin{abstract}
The recent Belle II observation of $B \to K E_{\rm miss}$ challenges theoretical interpretations in terms of Standard Model neutrino final states. Instead, we consider new physics scenarios where up to two new light-invisible particles of spin 0 up to 3/2 are present in the final state. We identify viable scenarios by reconstructing the (binned) likelihoods of the relevant $B \to K^{(*)} E_{\rm miss}$ and also $B_s \to E_{\rm miss}$ experimental analyses and present preferred regions of couplings and masses. In particular, we find that the current data prefers two-body decay kinematics involving the emission of a single massive scalar or a vector particle, or alternatively, three-body decays involving pairs of massive scalars or spin 1/2 fermions. When applicable, we compare our findings with existing literature and briefly discuss some model-building implications.
\end{abstract}

\maketitle

%%%%%%%%%%%%%%%%%%%%%%%%%%%%%%%%%%%%%%%%%%
%
\section{Introduction}
\label{sec:Intro}
%
%%%%%%%%%%%%%%%%%%%%%%%%%%%%%%%%%%%%%%%%%%

Historically, rare decays of (heavy) flavoured mesons have been important probes and harbingers of new physics (NP). In the last decade, the LHCb and B-factory experiments have produced several intriguing results on rare semileptonic $B$ meson decays, including the charged current mediated $B \to D^{(*)} \tau \nu$ and the flavour-changing neutral current (FCNC) mediated $B \to K^{(*)} \mu^+ \mu^-$ (see, e.g., Refs.~\cite{HFLAV:2022esi, Alguero:2023jeh}), that challenge explanations within the Standard Model (SM).

Most recently, the Belle~II experiment has measured the branching ratio $\mathcal{B}(B\to K E_{\text{miss}}) = (2.3 \pm 0.7) \times 10^{-5}$. 
Assuming that the missing energy $E_{\text{miss}}$ is carried away by a pair of massless SM neutrinos, the result lies $2.9 \sigma$ above the SM prediction for $\mathcal{B}(B\to K \nu \bar\nu)$~\cite{Belle-II:2023esi}. The quark-level transition $b \to s  \nu \bar \nu$ is also probed by the complementary mode $B \to K^{\ast} E_{\text{\rm miss}}$, with current experimental upper bounds on the branching ratio of the order $\mathcal{B}(B^{ } \to K^{*}E_{\text{\rm miss}}) < 11 \times 10^{-5} $~\cite{BaBar:2013npw}. { Finally, an  upper limit on the branching fraction of invisible $B_s$ decays, $\mathcal{B}(B_s \to E_{\rm miss}) < 5.4 \times 10^{-4}$ (90\% CL), has been recently derived in Ref.~\cite{Alonso-Alvarez:2023mgc} using a recast of ALEPH data~\cite{ALEPH:2000vvi}.} While innocuous in the SM, we show that this mode can put competitive bounds on particular NP interpretations of the Belle II result.

Undetected particles (neutrinos) in the final state make these FCNC processes experimentally more challenging compared to those producing charged leptons. On the other hand, they are theoretically cleaner~\cite{Buras:2020xsm,Buras:2014fpa}. In particular, the relevant hadronic matrix elements are well understood within existing theoretical frameworks~\cite{Becirevic:2023aov, Gubernari:2023puw,Athron:2023hmz}.
A variety of NP models aiming to resolve the Belle~II excess have been proposed in the literature~\cite{Bause:2023mfe,Allwicher:2023xba,Athron:2023hmz,Chen:2023wpb,Felkl:2023ayn,Abdughani:2023dlr,Dreiner:2023cms,He:2023bnk,Berezhnoy:2023rxx,Datta:2023iln,Altmannshofer:2023hkn,Fridell:2023ssf,Gabrielli:2024wys, Chen:2024jlj, Hou:2024vyw,Ho:2024cwk}.
In some scenarios, the SM neutrinos still carry away all of the observed missing energy~\cite{Bause:2023mfe,Allwicher:2023xba,Chen:2023wpb,Athron:2023hmz}, while in others, novel undetected decay products are also present in the final state~\cite{Altmannshofer:2023hkn,Fridell:2023ssf,Gabrielli:2024wys,Ho:2024cwk}. 
Restricting the outgoing invisible states to SM neutrinos only, there are two possibilities for the associated NP; either it couples universally to all three lepton generations~\cite{Athron:2023hmz, Allwicher:2023xba}, or it prefers some (e.g., $\nu_\tau$) neutrino flavours~\cite{Bause:2023mfe,Allwicher:2023xba},
leading to the violation of lepton flavour universality (LFU). A general analysis of LFU and non-LFU NP, coupling to both left- and right-handed quark current operators and contributing to the decays $B\to K^{(\ast)} \ell^{+} \ell^{-}$ and $B\to K^{(\ast)}  \nu \bar \nu$, was presented in Ref.~\cite{Descotes-Genon:2020buf}. It showed that enhancing the $B\to K \nu \bar \nu$ branching ratio while simultaneously satisfying existing constraints on $\mathcal{B} (B \to K^{*}\nu \bar{\nu})$ necessitates a significant contribution from right-handed quark current operators. In the LFU limit, however, these operators induce large contributions to the $C_{9,10}^{\prime}$ Wilson coefficients, present in the $b \to s \ell^{+} \ell^{-}$  effective Hamiltonian, which are already ruled out by existing measurements of $B\to K^{(*)} \mu^+ \mu^-$ and $B_s \to \mu^+ \mu^-$ decays~\cite{Alguero:2023jeh}.  
Thus, the only remaining phenomenologically viable option is LFU-violating NP, which couples both left- and right-handed quark currents predominantly to $\tau$ neutrinos~\cite{Bause:2023mfe,Allwicher:2023xba}.
On the other hand, the presence of additional invisible final states would circumvent the need for LFU-violating NP by decoupling the $B\to K^{} E_{\text{\rm miss}}$ measurement from the constrains  imposed by $b\to s \ell^+\ell^-$ transitions.

Motivated by the remarkable Belle~II measurement of $B\to K^{} E_{\text{\rm miss}}$ and the phenomenological difficulties in accommodating the observed excess exclusively with SM neutrinos, we turn to the interesting scenario that additional light invisible states are present in the final state. 
We systematically consider both single scalar and vector particle final states, as well as pairs of scalars, spin $1/2$ and $3/2$ fermions, and vectors, following 
Ref.~\cite{Kamenik:2011vy}. Since several of these possibilities significantly alter the phase space and kinematic distributions of events in the experiments, we consider not only the total branching ratios $\mathcal{B}(B \to K^{(*)} E_{\text{miss}})$, but also all available distributions presented in the Belle~II and BaBar analyses. {Finally, interactions producing two invisible particles in the final state of $B\to K^{}E_{\text{miss}}$ can induce the invisible $B_s$ decay $B_s \to E_{\text{miss}}$ whose larger phase space can probe massive invisible states beyond the kinematical limit of $B\to K^{*}E_{\text{miss}}$. The experimental upper bound from ALEPH~\cite{Alonso-Alvarez:2023mgc,ALEPH:2000vvi} can thus be considered as complementary.} 

Using this data, we construct our likelihoods for the possible invisible final states, discerning which scenarios are favoured (and, if so, what masses and couplings are implied).

We note that partial analyses of some of the scenarios considered in this work have already been performed in the literature~\cite{Felkl:2023ayn,Abdughani:2023dlr,Dreiner:2023cms,He:2023bnk,Berezhnoy:2023rxx,Datta:2023iln,Altmannshofer:2023hkn,Fridell:2023ssf}. However, a comprehensive study of all possibilities taking into account all available decay distributions has not been implemented to date. When applicable, we compare our results and findings with the previously published results.

\begin{figure}[t!]
  \centering
  \includegraphics[width=0.65 \columnwidth]{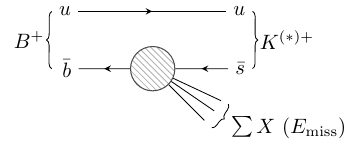}
  \includegraphics[width=0.58 \columnwidth]{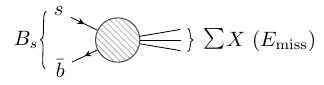}
  \caption{(Above) The process $B^+ \to K^{(*)+}\sum X$, resulting in the $B \to K E_{\text{miss}}$ signal. In the SM, $\sum X = \nu\bar{\nu}$, but additional light invisible particles (for example, a pair of heavy Majorana fermions, $\sum X = N\bar{N}$) may be present in the final state. (Below) The same invisible final states with the same couplings induce the decay $B_s \to \sum X$, leading to the $B_s \to E_{\rm miss}$ signal.}
  \label{fig:decay-diagram}
\end{figure}

The remainder of this work is structured as follows. In Section~\ref{sec:Model}, we introduce the NP fields describing the invisible final states, outlining their different (effective) couplings to the quark currents relevant to $B$ meson decays. We describe some features of the kinematic distributions of the decays, while explicit expressions for the branching ratios are given in Appendix~\ref{app:decays}. In Section~\ref{sec:Methods}, we give an overview of how Belle~II and BaBar data are used to construct likelihoods for the different NP scenarios.

In Section~\ref{sec:Results}, we then discuss the results of minimising these (negative) log-likelihoods as a function of the invisible particle masses and couplings. Firstly, we show that certain scenarios are immediately disfavoured. Of the scenarios that remain, we explore what masses and couplings are implied by the Belle~II excess and are at the same time compatible with existing constraints from BaBar and ALEPH. We also compare our findings to existing results in the literature and briefly discuss some model-building implications before concluding in Section~\ref{sec:conclusions}.
%%%%%%%%%%%%%%%%%%%%%%%%%%%%%%%%%%%%%%%%%%
%
\section{Model Considerations}
\label{sec:Model}
%
%%%%%%%%%%%%%%%%%%%%%%%%%%%%%%%%%%%%%%%%%%

As depicted in Fig.~\ref{fig:decay-diagram}, we consider the contribution of additional invisible final states, denoted as $\sum X$, to $B\to K^{(*)} E_{\text{miss}}$. This is alongside the SM neutrinos, $\sum X = \sum_\alpha \nu_\alpha\bar{\nu}_\alpha \equiv \nu\bar{\nu}$, with $\alpha \in \{e,\mu,\tau\}$. Any number of invisible final states may be present; however, to avoid phase space suppression we will consider only one or two invisible final state particles. With this requirement, there are now only a few possible scenarios. Considering fields $X\in\{\phi, \psi, V_\mu, \Psi_\mu\}$, corresponding to (massive) particles of spin $J = \{0,1/2,1,3/2\}$, respectively,
leads to the following possible final states:
\begin{align}
\sum X \in \{\phi, V, \phi\bar{\phi}, \psi\bar{\psi}, V\bar{V}, \Psi\bar{\Psi}\}\,.
\end{align}
Thus, two-body decays can only involve scalar and vector bosons, $B\to K^{(*)}\phi/V$. Three-body decays can proceed to pairs of scalars, vectors, and also spin $1/2$ and $3/2$ fermions, $B\to K^{(*)}\phi\bar{\phi}/\psi\bar{\psi}/V\bar{V}/\Psi\bar{\Psi}$.

We assume that the invisible state is a singlet under the SM gauge group $SU(3)_c \times SU(2)_L \times U(1)_Y$, but leave open the possibility that it is charged under a dark gauge or global symmetry. As a result, any observable effect of the invisible states or hidden sector must come from interactions involving gauge-invariant combinations of SM fields. Of these interactions, those which are renormalisable, i.e. dimension-four operators, are sometimes referred to as \textit{portals}. For example, the generic fermion field $\psi$ may correspond to a right-handed gauge singlet fermion $N_R$, with the well-known mixing portal $-Y_\nu \bar{l}\tilde{H}N_R$ and also a Majorana mass term $-\frac{1}{2}M_R\bar{N}_R^c N_R$ if lepton number is not conserved. These terms result in the Type-I seesaw mechanism, generating masses for the light neutrinos. A generic vector field $V_\mu$ may instead exhibit kinetic mixing with the SM hypercharge, e.g. $-\frac{\varepsilon}{4}B_{\mu\nu} V^{\mu\nu}$, with $V_{\mu\nu} = \partial_{\mu}V_\nu - \partial_{\nu}V_\mu$. 
{Recently~\cite{Dror:2017nsg, DiLuzio:2022ziu} it has been shown new massive vector fields may also couple directly to  SM currents as $\mathcal J_\mu V^\mu$, even if $\mathcal J$ is not conserved.}
Finally, the so-called Higgs portal, with the terms $\mu' (H^\dagger H)\phi$ and $\lambda' (H^\dagger H)\phi^\dagger\phi$, is another possible window to NP in the form of a scalar $\phi$.

However, heavy NP may also mediate the interactions between new light-invisible particles and the SM. At energies below the associated NP scale $\Lambda$, these then manifest as higher-dimensional effective operators, i.e.
\begin{align}
\mathcal{L} = \mathcal{L}_{\text{SM}+X} + \sum_i C_{i}^{(d)}\mathcal{O}_{i}^{(d)}\,.
\end{align}
Here, $\mathcal{L}_{\text{SM}+X}$ is the dimension-four SM Lagrangian extended with the gauge singlet field(s) $X$, while the sum denotes a tower of dimension-$d$ SM gauge-invariant operators $\mathcal{O}_{i}^{(d)}$. In the standard dimensional analysis, the Wilson coefficients $C_i^{(d)}$ are proportional to $4-d$ powers of the heavy NP scale $\Lambda$.

At energies below the weak scale, the $X$-extended SM gauge-invariant effective field theory 
is no longer appropriate. Instead, the weak effective theory in the broken phase of the SM, i.e. operators invariant under $SU(3)_c \times U(1)_{Q}$, should be used. 
Since in this EFT, the left- and right-handed chiral fermion fields carry identical charges,
we can consider the parity basis for the quark fields, i.e. either $P$-even or $P$-odd quark currents. In the context of $B$ meson decays, the expressions for the branching ratios are simplified considerably. 

Using the parity basis for $b\to s$ transition quark currents, the interactions of the generic invisible fields $X \in \{\phi,\psi,V_\mu,\Psi_\mu\}$ in the weak effective theory (including operators up to dimension $d=6$) are as follows. Starting with the $P$-even quark currents, the coupling of the vector quark current to invisible states is described by the effective Hamiltonian
\begin{align}
\label{eq:L_V}
\mathcal{H}_{\text{eff}}^{V} & \supset \bar{s} \gamma_\mu b\bigg[h_V V^\mu + \frac{g_{VV}}{\Lambda^2} i \phi^\dagger \overset{\leftrightarrow~\,}{\partial^\mu} \phi \nonumber \\
&\hspace{3.6em} + \frac{f_{VV}}{\Lambda^2}\bar{\psi}\gamma^\mu \psi + \frac{f_{VA}}{\Lambda^2}\bar{\psi}\gamma^\mu \gamma_5 \psi \nonumber \\
&\hspace{3.6em}  + \frac{F_{VV}}{\Lambda^2}\bar{\Psi}^\rho \gamma^\mu \Psi_\rho + \frac{F_{VA}}{\Lambda^2}\bar{\Psi}^\rho \gamma^\mu \gamma_5 \Psi_\rho\bigg]\,,
\end{align}
where the corresponding hermitian conjugate terms are implicit in all our Hamiltonians. Note that we have omitted the operator $g_V(\bar{s}\gamma_\mu b)\partial^\mu \phi/\Lambda$, because it can be rewritten using the quark field equations of motion as $g_S(\bar{s}b)\phi$, with $g_S = i m_b g_V/\Lambda$. This leads us to the general effective Hamiltonian of the scalar quark current,
\begin{align}
\label{eq:L_S}
\mathcal{H}_{\text{eff}}^{S} &\supset \bar{s} b\bigg[g_S \phi + \frac{g_{SS}}{\Lambda} \phi^\dagger\phi + \frac{h_{S}}{\Lambda}V_{\mu}^\dagger V^{\mu} \nonumber\\
&\hspace{2.7em} + \frac{f_{SS}}{\Lambda^2}\bar{\psi}\psi + \frac{f_{SP}}{\Lambda^2}\bar{\psi}\gamma_5\psi \nonumber\\
&\hspace{2.7em} + \frac{F_{SS}}{\Lambda^2}\bar{\Psi}^\rho \Psi_\rho + \frac{F_{SP}}{\Lambda^2}\bar{\Psi}^\rho \gamma_5 \Psi_\rho \bigg]\,.
\end{align}
Finally, the effective Hamiltonian for the tensor quark current is
\begin{align}
\label{eq:L_T}
\mathcal{H}_{\text{eff}}^{T} & \supset \bar{s} \sigma_{\mu\nu} b\bigg[\frac{h_T}{\Lambda}V^{\mu\nu} + \frac{f_{TT}}{\Lambda^2}\bar{\psi}\sigma^{\mu\nu}\psi + \frac{F_{TT}}{\Lambda^2}\bar{\Psi}^\rho \sigma^{\mu\nu} \Psi_\rho  \nonumber \\
& \hspace{4em} + \frac{F_{TS}}{\Lambda^2}\bar{\Psi}^{[\mu} \Psi^{\nu]} + \frac{F_{TP}}{\Lambda^2}\bar{\Psi}^{[\mu} \gamma_5\Psi^{\nu]} 
\bigg] \,,
\end{align}
with $\bar{\Psi}^{[\mu} \Gamma \Psi^{\nu]} = i(\bar{\Psi}^{\mu} \Gamma \Psi^{\nu}-\bar{\Psi}^{\nu} \Gamma \Psi^{\mu})/2$. For the $P$-odd quark currents, the effective Hamiltonians are similar to the expressions above. In particular, for the axial vector quark current, the couplings to invisible states can be found by replacing $\bar{s}\gamma_\mu b \to \bar{s}\gamma_\mu\gamma_5 b$ and $V \to A$ in Eq.~\eqref{eq:L_V}, e.g. $h_V \to h_A$ and $f_{VV} \to f_{AV}$. Likewise, the couplings for the pseudoscalar and axial tensor quark currents are found with the replacements $\bar{s}b\to \bar{s}\gamma_5 b$ ($S \to P$) and $\bar{s}\sigma_{\mu\nu}b\to \bar{s}\sigma_{\mu\nu}\gamma_5 b$ ($T\to \tilde{T}$) in Eqs.~\eqref{eq:L_S} and~\eqref{eq:L_T}, respectively.

For completeness, in Appendix~\ref{app:basis}, we give the matching relations between the coefficients of the weak EFT operators in Eqs.~(\ref{eq:L_V}--\ref{eq:L_T}), written in the parity basis, and the coefficients of SM gauge-invariant operators, necessarily in the chiral basis for the quark fields, as defined in Ref.~\cite{Kamenik:2011vy}. For example, the scalar and pseudoscalar operators $g_S (\bar{s}b)\phi$ and $g_P(\bar{s}\gamma_5 b)\phi$ are induced by the SM gauge-invariant operators $C_{d\phi}^{S,L}  \bar{d}_R H^\dagger q \phi/\Lambda$ and $C_{d\phi}^{S,R} \bar{q} H d_R \phi/\Lambda$, with the matching $g_{S(P)} = v/\sqrt{2}\Lambda \times (C_{d\phi}^{S,R}\pm C_{d\phi}^{S,L})/2$.

We now comment on some interesting formal details regarding the possible invisible final states. If the generic scalar $\phi$ or vector $V_\mu$ is charged under a dark, possibly non-Abelian, gauge group, the operators $g_{S(P)} [\bar{s}(\gamma_5)b] \phi$ and $h_{V(A)} [\bar{s}\gamma_\mu (\gamma_5) b] V^\mu$ must vanish and it is forbidden to emit a single or vector scalar boson. If the scalar field is neutral, i.e., a real scalar field, the operator $g_{VV} (\bar{s}\gamma_\mu b)i\phi^* \overset{\leftrightarrow~\,}{\partial^\mu} \phi/\Lambda^2$ vanishes instead.

\begin{figure}[t!]
  \centering
  \includegraphics[width=0.9\columnwidth]{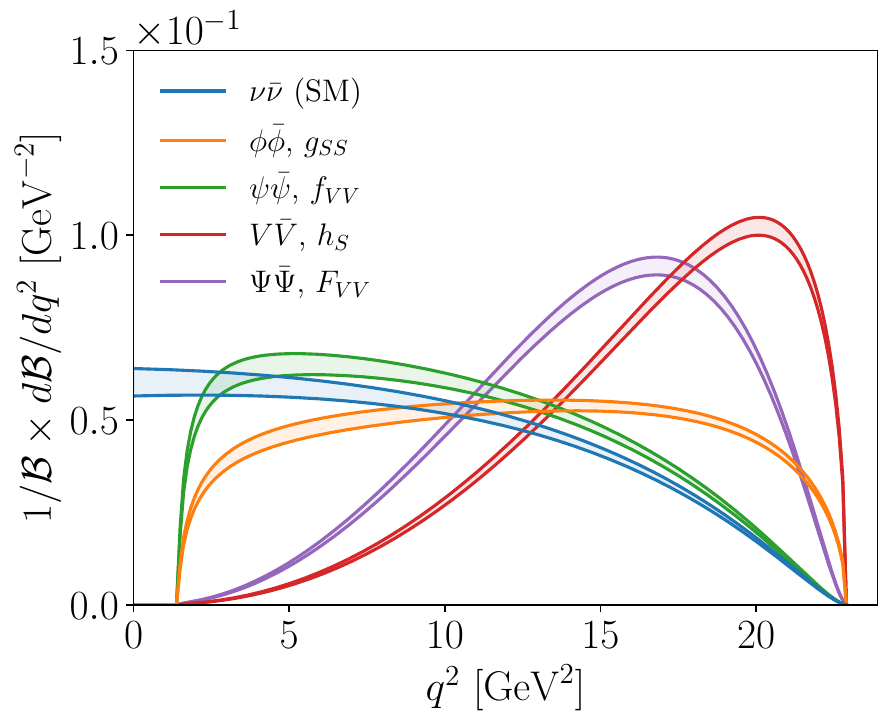}
  \caption{Normalised differential branching fraction of the three-body decay $B\to K \sum X$ as a function of the momentum transfer $q^2$ for SM neutrinos and additional NP light-states, $\sum X= \nu\bar\nu,\,\phi\bar\phi,\,\psi\bar\psi,\,V\bar{V},\,\Psi\bar\Psi $.}
  \label{fig:diff-BRs}
\end{figure}

The generic fermion field $\psi$, if massive, may either be a Dirac or Majorana fermion. In the latter scenario ($\psi = \psi^c$), only viable if $\psi$ is a singlet under a dark gauge group, the vector and tensor fermion bilinears vanish identically, $\bar{\psi}\gamma_\mu\psi = \bar{\psi}\sigma_{\mu\nu}\psi = 0$. 
For the other fermion bilinears, two possible contractions are possible for the Majorana fields, so the amplitudes for Majorana fermions are a factor of two larger than those for Dirac fermions. On the other hand, the final phase space integration has to take two identical particles into account, reducing the rate by a factor of two. Similar observations apply to all scenarios with two identical particles in the final state (if $\phi= \phi^\dagger $, $V= V^\dagger$ or $\Psi=\Psi^c$).

Finally, decay rates derived from the operators $h_{V(A)} [\bar{s}\gamma_\mu (\gamma_5) b] V^\mu$ and $h_{S(P)} [\bar{s}(\gamma_5)b] V_{\mu}^\dagger V^{\mu}/\Lambda$ diverge in the limit where the generic vector field $V_\mu$ is massless, $m_V \to 0$. This is an expected behaviour because the massless limit can only be consistently defined when $V_\mu$ is a gauge field. It could be the case that $V_\mu$ gets its mass from a dark Higgs-like mechanism, such that $m_V$ is proportional to some power of the couplings $h_{V(A)}$ multiplied by the vacuum expectation value of the dark Higgs field. Then, the massless limit implies that the couplings $h_{V(A)}$ vanish, avoiding the divergence. Alternatively, operators can be constructed solely from the manifestly dark gauge-invariant field strength tensor $V_{\mu\nu}$. Then, the operators with couplings $h_V$ and $h_A$ are generated by applying the equations of motion for $V_\mu$ to such gauge-invariant operators, i.e. $(\bar{s}\gamma_\mu b)\partial_\nu V^{\mu\nu}/\Lambda^2 \to m_V^2 (\bar{s}\gamma_\mu b) V^\mu/\Lambda^2$, and consequently the $1/m_V^2$ divergences in decay rates are tamed via the replacement $ 1/m_V^2 \to 1/\Lambda^2$. The situation is analogous for the spin $3/2$ fermions.

With these details in mind, we derive the branching ratios for the two-body processes $B\to K^{(*)}\phi/V$ and differential branching ratios in the momentum transfer $q^2$ for the three-body processes $B\to K^{(*)}\phi\bar{\phi}/\psi\bar{\psi}/V\bar{V}/\Psi\bar{\Psi}$, given in full in Appendix~\ref{app:decays}. We do not list the differential branching ratios for $B\to K^{*}V\bar{V}/\Psi\bar{\Psi}$, because we find that non-zero values of the couplings $h_{S,P,T}$ and $F_{XY}$ are not favoured by the Belle~II results~\cite{Belle-II:2023esi}. This result is foreshadowed by Fig.~\ref{fig:diff-BRs}, where we display the normalised differential branching fractions for the three-body decays, for the SM neutrinos $\sum X = \nu\bar{\nu}$ (blue) and the light NP states $\sum X = \phi\bar{\phi}, \psi\bar{\psi}, V\bar{V}, \Psi\bar{\Psi}$ (orange, green, red and purple, respectively), with the choice $m_X = 0.6$~GeV. The bands illustrate the theoretical uncertainties from the $B\to K$ form factors, implemented as outlined in Appendix~\ref{app:ffs}. It can be seen that the differential $B\to K\psi\bar{\psi}$ rate peaks close to the $\psi\bar{\psi}$ threshold, while $B\to KV\bar{V}$ and $B\to K\Psi\bar{\Psi}$ peak closer to the endpoint. The distribution of the scalar final states exhibits a relatively flat $q^2$ distribution. Since most of the signal observed by Belle II peaks at low $q^2$, see Fig.~\ref{fig:events}, this generically disfavours the three-body decay scenarios with spin 1 and 3/2 final states. 

%%%%%%%%%%%%%%%%%%%%%%%%%%%%%%%%%%%%%%%%%%
%
\section{Methodology}
\label{sec:Methods}
%
%%%%%%%%%%%%%%%%%%%%%%%%%%%%%%%%%%%%%%%%%%

Next, we summarise how a possible NP contribution to the $B \to K^{(*)}E_{\text{miss}}$ decay translates to a signal in the Belle~II and BaBar experiments. We also describe how the expected SM plus NP signal is used to construct likelihoods for the different NP scenarios, given the distribution of events seen in each experiment. Finally, we briefly describe the recast of the upper limit on $B_s \to E_{\rm miss}$ decays from ALEPH~\cite{Alonso-Alvarez:2023mgc,ALEPH:2000vvi}.

\begin{figure}
  \centering
  \includegraphics[width=0.9\columnwidth]{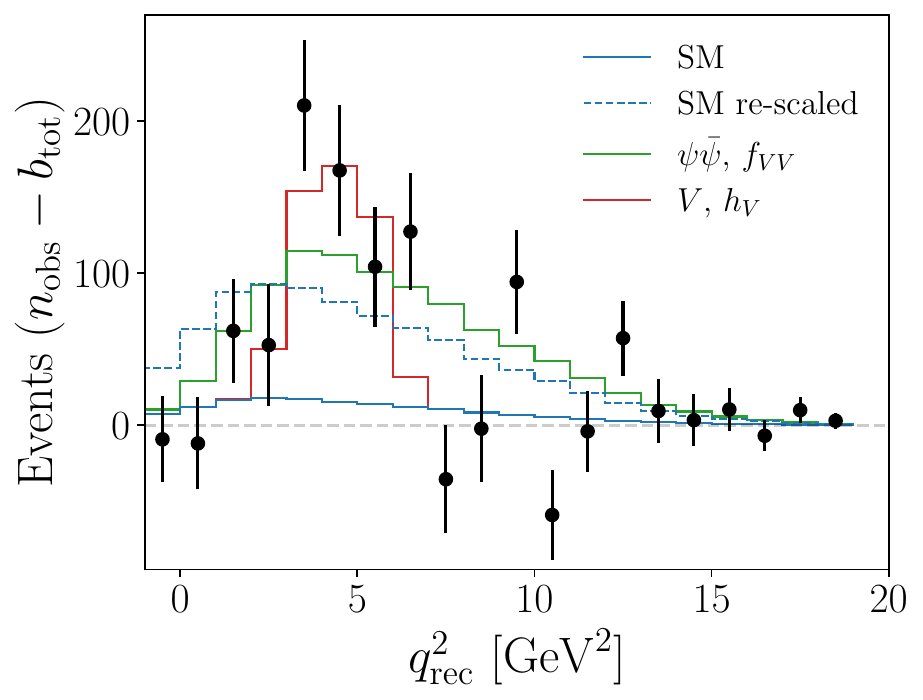}
  \caption{The number of $B^+ \to K^+ E_{\text{miss}}$ events in $q_{\text{rec}}^2$ in the $\eta(\text{BDT}_2)> 0.92$ Belle~II signal region, with the total background subtracted (black dots, with error bars showing the total statistical uncertainty). Shown for comparison is the predicted SM distribution, before (blue, solid) and after (blue, dashed) re-scaling by the best-fit signal strength, and also after adding the distribution for $B^+\to K^+\bar{\psi}\psi$ (green) for $m_\psi = 0.6~\text{GeV}$ and $f_{VV}/\Lambda^2 = 1.7\times 10^{-2}~\text{TeV}^{-2}$ and for $B^+\to K^+ V$ (red) for $m_V = 2.1~\text{GeV}$ and $h_V = 7.1 \times 10^{-9}$.}
  \label{fig:events}
\end{figure}

The Belle~II experiment conducted a search for the SM process $B^+\to K^+\nu\bar{\nu}$ using $e^+e^- \to B^+B^-$ at the $\Upsilon(4S)$ resonance with an integrated luminosity of $\mathcal{L} = 362 ~\text{fb}^{-1}$~\cite{Belle-II:2023esi}. The collaboration used two methods: an inclusive (ITA) and hadronic (HTA) tagging analysis. The former exploits inclusive properties of the $B^+B^-$ pair, while the latter uses an explicit reconstruction of the partner $B$ meson via its hadronic decay. Thus, the former method trades a higher signal efficiency for an increased background. The known backgrounds, separated as $B^+B^-$, $B^0\bar{B}^0$ and continuum components, were suppressed by utilising the kinematic properties of the decay in a multivariate classifier, i.e. training a boosted decision tree (BDT). Cuts are then placed on the classifier used for final event selection, $\text{BDT}_2$, to optimise the signal region.
Using the lower purity signal region, $\eta(\text{BDT}_2) > 0.92$, where $\eta(\text{BDT}_2)$ is defined in Eq.~(4) of Ref.~\cite{Belle-II:2023esi}, Belle~II observed a signal corresponding to a branching ratio $2.9 \sigma$ or a factor of four above the SM prediction. The observed events minus the backgrounds are shown in bins of the reconstructed momentum transfer, $q_{\text{rec}}^2$, in Fig.~\ref{fig:events}.

The BaBar experiment has also performed a search for the $B\to K^{(*)}\nu\bar{\nu}$ process~\cite{BaBar:2013npw}. Using $e^+e^-$ collisions, again at the $\Upsilon(4S)$ resonance and with an integrated luminosity of $\mathcal{L} = 429~\text{fb}^{-1}$, BaBar used the HTA method to search for both neutral and charged invisible $B$ decays. Using cuts to define a signal region minimises the $B^{+} B^{-}$ and continuum backgrounds, separated as those which are correctly reconstructed (peak) and those which are not (combinatorial). Their results were consistent with the SM, and no significant signal was observed, resulting in upper bounds between six and twelve times the SM rates.

For the SM or invisible final state(s), $\sum X$, we determine the distribution of Belle~II and BaBar events in the reconstructed momentum transfer, $q_{\text{rec}}^2$, as
\begin{align}
\label{eq:exp_sig}
\frac{dN_{\text{SM}(X)}}{dq^2_{\mathrm{rec}}} = N_B \int dq^2 f_{q^2_\mathrm{rec}}(q^2) \epsilon(q^2)\frac{d\mc{B}_{\text{SM}(X)}}{dq^2}\,,
\end{align}
where $N_B$ is the number of $B^{+}B^{-}$ or $B^{0}\bar{B}^{0}$ pairs, $f_{q^2_\mathrm{rec}}(q^2)$ corresponds to the smearing of $q_{\text{rec}}^2$ with respect to the true momentum transfer $q^2$, and $\epsilon(q^2)$ is the detector efficiency as a function of momentum transfer. The kinematic range is $(\sum m_X)^2 < q^2 < (m_B - m_{K^{(*)}})^2$. We provide details on how we obtained $f_{q^2_\mathrm{rec}}(q^2)$ and $ \epsilon(q^2)$ for the Belle II and BaBar analyses in Appendix~\ref{app:ExpLikelihoodReconstruction}.

\begin{figure}[t!]
  \centering
  \includegraphics[width=0.9\columnwidth]{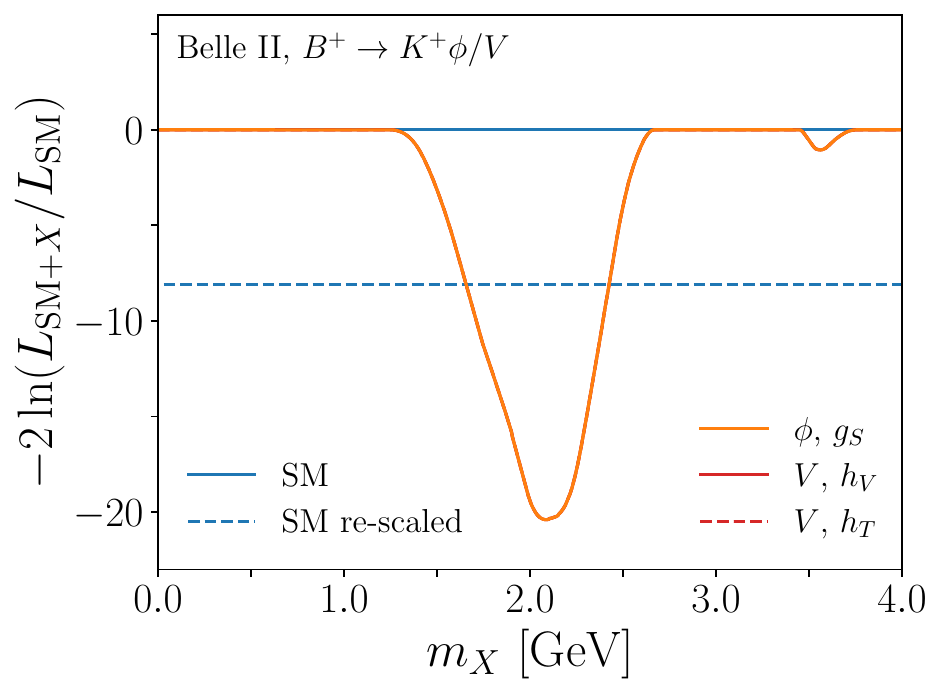}
  \caption{Minimised log-likelihood ratio of the SM plus $\phi/V$ scenario with respect to the SM-only hypothesis as a function of the scalar/vector mass $m_{\phi/V}$. The log-likelihood ratio $t_{\phi/V}$ for single effective coupling contributions $g_S$, $h_V$ and $h_T/\Lambda$ to $B\to K \phi/V$ is minimised over the coupling and nuisance parameters $(\boldsymbol{\theta}_x,\tau_b)$ for each $m_{\phi/V}$. As the likelihood is independent of the nature of the light NP state and the coupling in two-body decays, the three lines overlap. The blue dashed line corresponds to the minimum of the log-likelihood ratio of the re-scaled SM hypothesis.}
  \label{fig:two-body-fits}
\end{figure}

We obtain the expected SM ($X$) signal $s_{\text{SM} (X)}^i$ for the bin $i$, $[q^2_{\mathrm{rec},i},q^2_{\mathrm{rec},i+1}]$, as 
\begin{align}
\label{eq:signal_in_bin}
s^i_{\text{SM}(X)} = \int_{q^2_{\mathrm{rec},i}}^{q^2_{\mathrm{rec},i+1}}dq^2_{\mathrm{rec}}\,\frac{dN_{\text{SM}(X)}}{dq^2_{\mathrm{rec}}}\,,
\end{align}
and the total expected event count in bin $i$ is,
\begin{align}
n_{\text{exp}}^i & = \mu \left(1 + \theta_{\text{SM}}^i\right) s_{\text{SM}}^i + \left(1 + \theta_{X}^i\right)s_{X}^i(m_X, c_X) \nonumber\\
& \quad + \sum_b \tau_b (1 + \theta_{b}^i) b^i\,,
\end{align}
where $\mu$ is a signal strength parameter allowing to re-scale the SM signal $s_{\text{SM}}^i$, the NP signal $s_{X}^i$ depends on the invisible particle mass $m_X$ and coupling $c_X$, $b^i$ is the binned expected signal for the background $b$, and $\tau_b$ is an overall normalisation. We account for systematic and Monte-Carlo statistical uncertainties via the nuisance parameters $\theta_x$. 

\begin{figure}[t!]
  \centering
  \includegraphics[width=0.9\columnwidth]{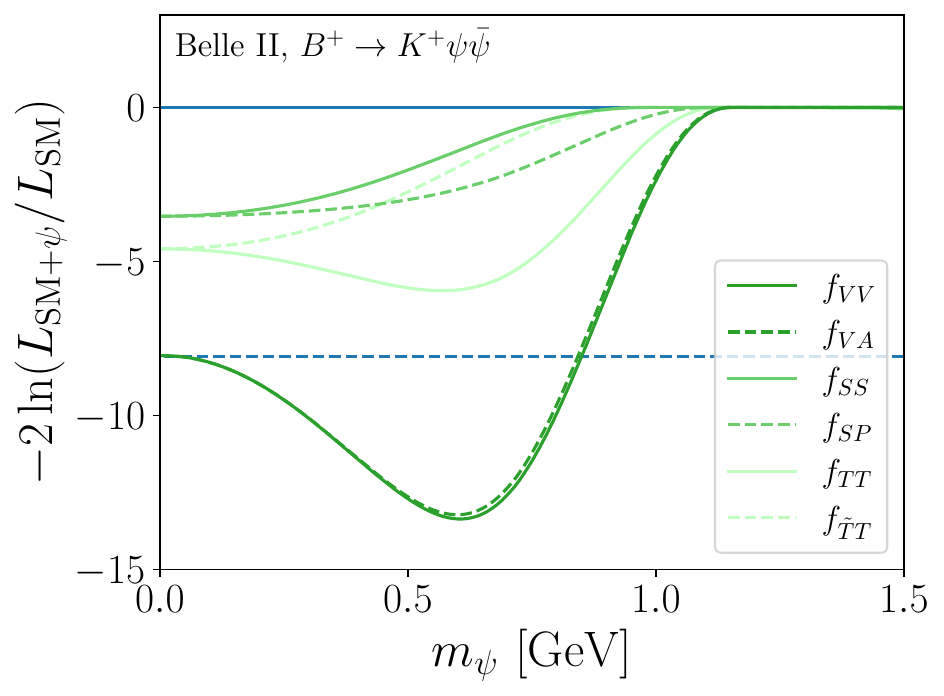}
  \includegraphics[width=0.9\columnwidth]{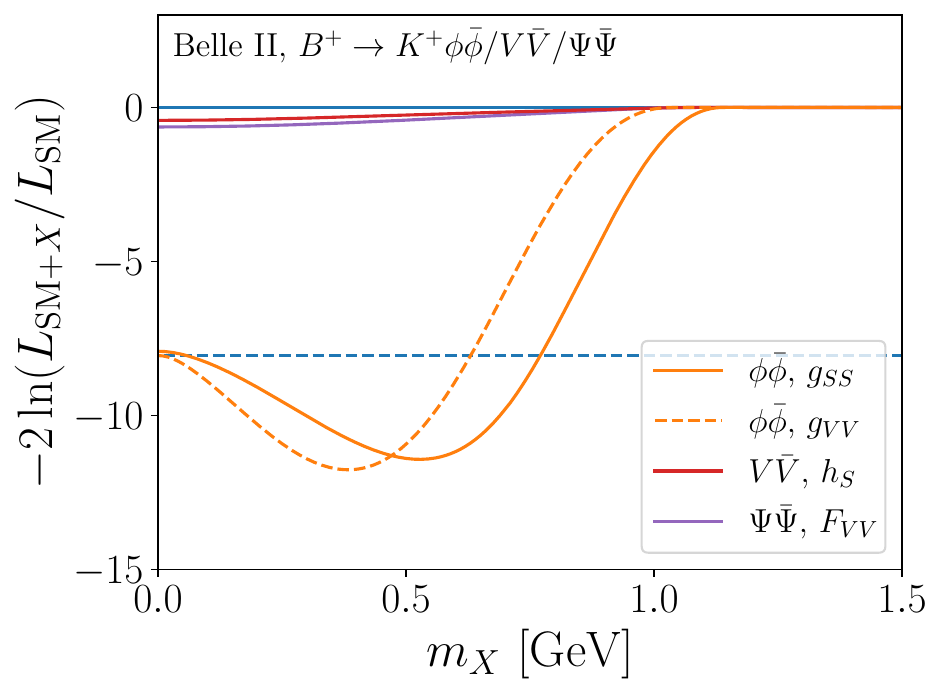}
  \caption{Minimised log-likelihood ratios of the SM plus three-body $\sum X$ scenarios with respect to the SM-only hypothesis as a function of the new particle $X$ mass $m_X$ for $\sum X=\psi\bar{\psi}$ (top) and $\sum X=\phi\bar{\phi}/V\bar{V}$ (bottom). The log-likelihood ratio $t_X$ for single effective coupling contributions $c_X$ to $B\to K \sum X$ is minimised over the coupling and nuisance parameters $(\boldsymbol{\theta}_x,\tau_b)$ for each $m_X$. The blue dotted line again corresponds to the re-scaled SM hypothesis.}
  \label{fig:three-body-fits}
\end{figure}

The combined likelihood then takes the following form
\begin{align}
L_{\text{SM}+X} &= \prod_i^{N_{\text{bins}}}\text{Poiss}\left[n_{\text{obs}}^i, n_{\text{exp}}^i(\mu, m_X, c_X, \boldsymbol{\theta}_x, \tau_b)\right]\nonumber\\
&\hspace{-0.5em}\quad \times \prod_{x = {\rm SM}, X, b}\mc{N}\left(\boldsymbol{\theta}_x; \boldsymbol{0}, \Sigma_x\right) \prod_{b} \mathcal{N}\left(\tau_b; 0, \sigma_b^2\right)\,,
\end{align}
where the event counts in each bin $n_i$ are Poissonian (Poiss) distributed, while we sample the nuisance parameters $\boldsymbol{\theta}_x$ from a multi-normal ($\cal N$) distribution centred at $\boldsymbol{0}$ and of covariance $\Sigma_x$. Finally, the overall normalisations of the backgrounds are sampled from a univariate normal distribution, centred at $0$ and with standard deviation $\sigma_b$.

The covariance for the SM signal, $\Sigma_{\text{SM}}$, is found by performing a Monte-Carlo simulation of the SM signal, $s_{\text{SM}}^i$, including the uncertainties on the efficiency and $B\to K^{(*)}$ form factors in Eq.~\eqref{eq:signal_in_bin}. The $q^2$ smearing introduces correlations among the $q^2_{\text{rec}}$ bins. The covariances for the background components, separated in the Belle~II and BaBar analyses as discussed above Eq.~\eqref{eq:signal_in_bin}, are found by simply re-scaling $\Sigma_{\text{SM}}$ to the relative size of the background. Finally, for the covariance for the NP signal, we take $(\Sigma_{X})_{ij} = s_{X}^i\delta_{ij}$, i.e. Poissonian uncertainties and neglecting correlations between bins. This speeds up considerably the following analysis, and we find that including the correlations has a negligible impact on the results.

In the following, we consider three types of signal hypothesis, corresponding to: (i) the SM-only scenario, where $\mu = 1$ and $s_X^i = 0$, (ii) the re-scaled SM scenario, where $\mu$ is treated as free nuisance parameter and $s_X^i = 0$, and lastly (iii) the SM plus various NP scenarios, with $\mu = 1$ and  $s_X^i\neq 0$, considering separately each NP final state $\sum X$ and its possible couplings $c_X$. On the one hand, the first two hypotheses serve to cross-check the validity of our likelihood recast, see Appendix~\ref{app:ExpLikelihoodReconstruction} for details. On the other hand, they represent important likelihood benchmarks against which we compare all of the NP scenarios. 

\begin{figure}[t!]
  \centering
  \includegraphics[width=0.7\columnwidth]{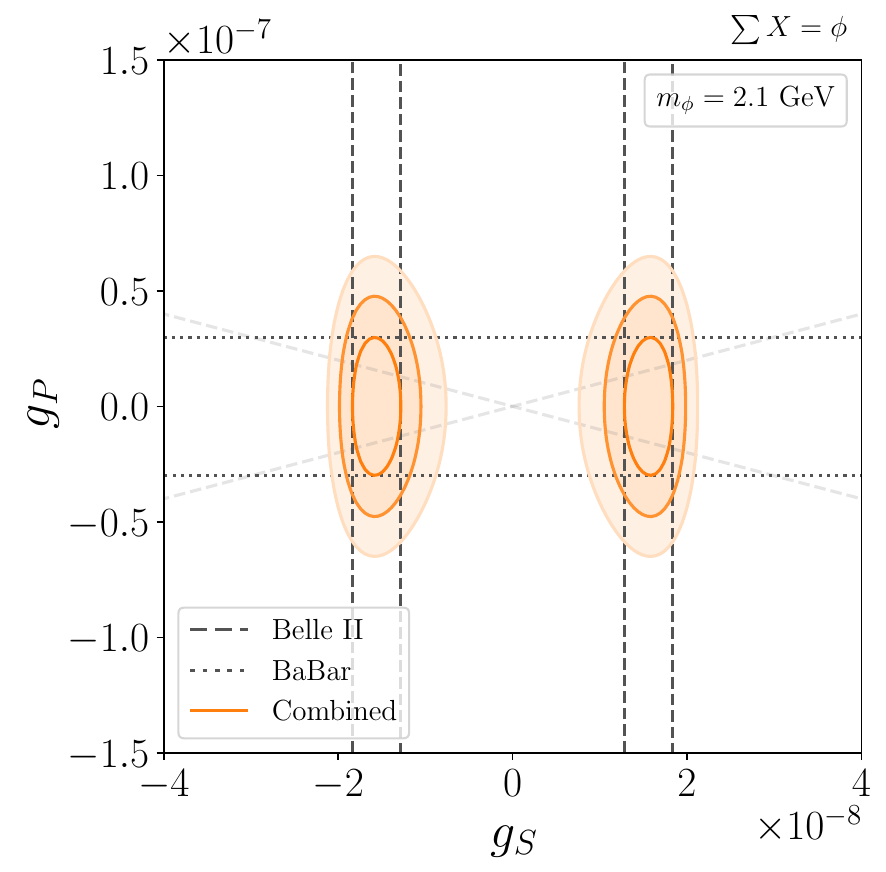}
  \caption{Favoured regions in the parameter space $(g_S, g_P)$ relevant to the two-body decay $B\to K^{(*)}\phi$, with the mass of $\phi$ fixed to a value close to the best-fit point, $m_\phi = 2.1$~GeV. Dashed and dotted grey lines correspond to 1-$\sigma$ confidence regions for the Belle II and BaBar analyses, respectively. Solid dark, medium and light orange lines correspond to the 1-, 2- and 3-$\sigma$ combined confidence regions. Diagonal grey dashed lines denote the chiral combinations of couplings $g_S = \pm g_P$.}
  \label{fig:two-body-scalar-coupling-constraints}
\end{figure}

For the SM + NP hypothesis, we define for convenience the log-likelihood ratio
\begin{align}
\label{eq:LL_ratio}
t_{X} = - 2 \ln \frac{L_{\text{SM}+X}}{L_{\text{SM}}}\,,
\end{align}
where $L_{\text{SM}}$ is the likelihood for the SM-only hypothesis. In the next section, we will use $t_X|_{\text{min}}$ (in which $L_{\text{SM}+X}$ and $L_{\text{SM}}$ are minimised with respect to the nuisance parameters $(\boldsymbol{\theta}_x,\tau_b)$ and the NP couplings or $\mu$, respectively) to see which invisible final states provide a better fit to the Belle~II excess compared to the (re-scaled) SM. We then examine the profile likelihoods $\hat{t}_X = t_X - t_X|_{\text{min}}$, where $t_X$ is not minimised with respect to the  NP couplings, to infer what NP couplings are implied by the data.

Lastly, we take into account the recent recast~\cite{Alonso-Alvarez:2023mgc}, of the ALEPH search for $b\to \tau^{-}\bar{\nu}_\tau X$ at LEP~\cite{ALEPH:2000vvi} in terms of an upper limit on invisible $B_s$ decays, $\mathcal{B}(B_s \to E_{\rm miss}) < 5.4 \times 10^{-4}$ (90\% CL). For $\sum X = \phi\bar{\phi}/\psi\bar{\psi}/V\bar{V}/\Psi\bar{\Psi}$ the decays $B_s \to \sum X$ depend on some of the same couplings as the $B\to K^{(*)}\sum X$ processes. We thus include this constraint in the likelihood for these scenarios as follows. Ref.~\cite{Alonso-Alvarez:2023mgc} provides the upper limit on the number of signal events at two confidence levels; from this information, the mean and standard deviation of the signal and thus branching fraction can be determined, which in turn can be used to construct a simple Gaussian log-likelihood with the NP prediction for the branching ratio, given in Appendix~\ref{app:decays}.

%%%%%%%%%%%%%%%%%%%%%%%%%%%%%%%%%%%%%%%%%%
%
\section{Results}
\label{sec:Results}
%
%%%%%%%%%%%%%%%%%%%%%%%%%%%%%%%%%%%%%%%%%%

%
\begin{figure}[t!]
  \centering
  \includegraphics[width=\columnwidth]{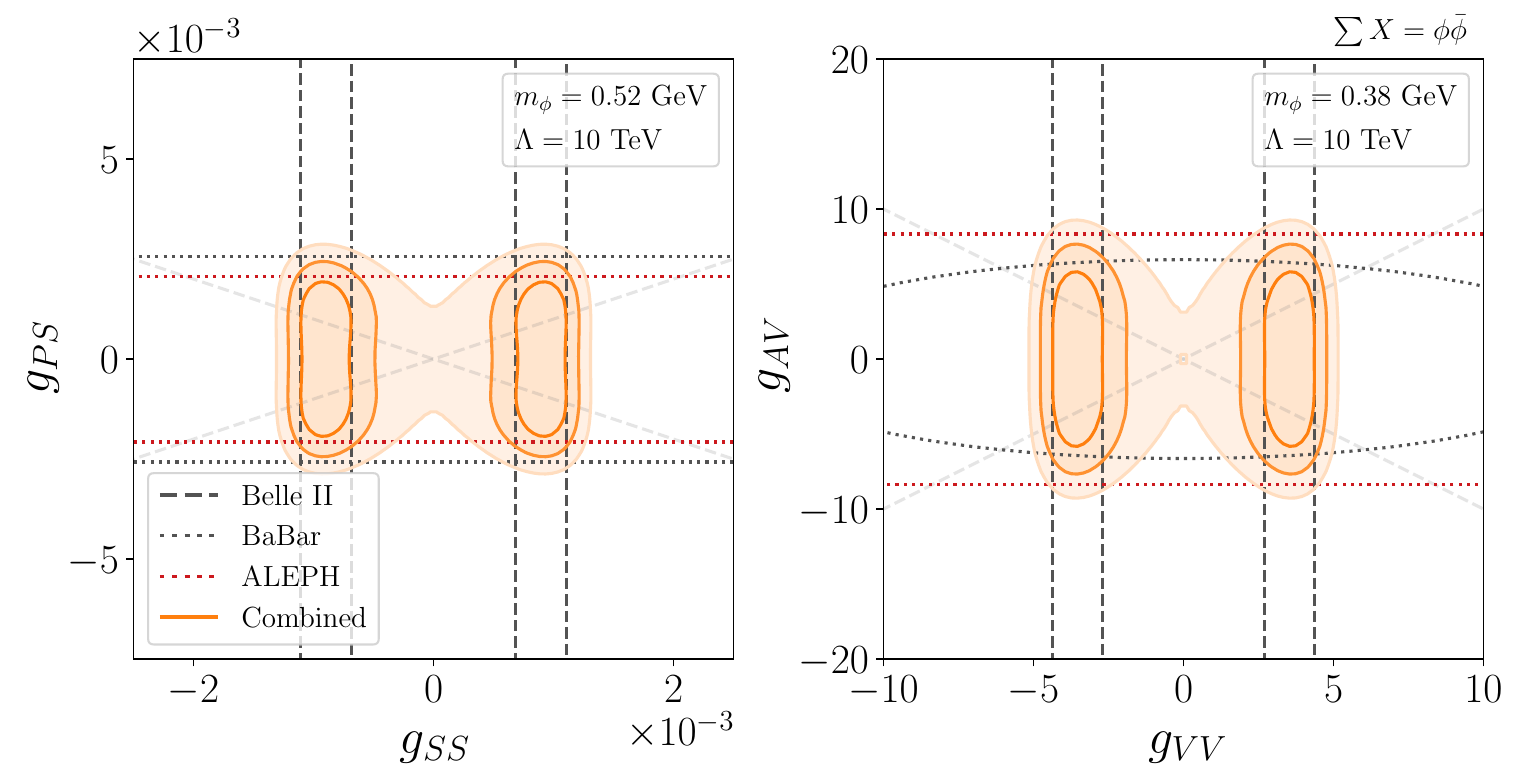}
  \caption{Favoured regions in the parameter space $(g_{PS}, g_{SS})$ (left) and $(g_{VV}, g_{AV})$ (right) relevant to the three-body decay $B\to K^{(*)}\phi\bar\phi$, with the mass of $\phi$ fixed to a value close to the respective best-fit point, $m_\phi = 0.52$~GeV and $m_\phi = 0.38$~GeV, and the heavy NP scale fixed at $\Lambda=10$~TeV. Colour coding is the same as Fig.~\ref{fig:two-body-scalar-coupling-constraints}. The red dashed lines indicate the 1-$\sigma$ confidence regions from ALEPH.
  }
  \label{fig:three-body-scalar-coupling-constraints}
\end{figure}

In the following, we present the results of minimising the binned log-likelihood ratio $t_X$, defined in Eq.~\eqref{eq:LL_ratio}, with respect to the model and nuisance parameters. To gain a better understanding of which NP scenarios are favoured by the excess seen by Belle~II, in each case, we find $t_X|_{\text{min}}$ for different values of the light invisible particle masses, $m_X$, using only the Belle~II data. For the scenarios that provide a better fit to the Belle~II data compared to the re-scaled SM prediction, we next profile over $t_X$ in the parameter space of effective couplings $c_X$, also including the BaBar and ALEPH data. In this way, we identify ranges of couplings compatible with both the Belle~II excess as well as the BaBar and ALEPH upper limits on $\mathcal{B}(B\to K^{*} E_{\text{miss}})$ and $\mathcal{B}(B_s\to E_{\rm miss})$, respectively. In the following, all minimisation is performed using the iminuit interface to the \textsc{minuit2} package~\cite{iminuit, James:1975dr}.

In Fig.~\ref{fig:two-body-fits}, we first show the outcome of minimising the binned likelihood ratio $t_{\phi/V}$ for the two-body decay scenarios $\sum X = \phi/V$. For different values of the masses $m_{\phi/V}$, the minimisation is performed with respect to the nuisance parameters~($\boldsymbol{\theta}_x,\tau_b)$ and the NP couplings; $g_S$ for the scalar boson and $h_V$ or $h_T/\Lambda$ for the vector boson. Firstly, we see that the minimised likelihoods in the three scenarios overlap because the likelihood is independent of the spin and coupling of single-particle $\sum X$; the likelihood is constructed in bins of $q_{\text{rec}}^2$, while the two-body scenario only depends on $q_{\text{rec}}^{2}$ through the smearing at the true momentum transfer $q^2 = m_{\phi/V}^2$. For masses in the range $1.7~\text{GeV} \lesssim m_{\phi/V} \lesssim 2.4~\text{GeV}$, the SM plus an invisible scalar or vector provides a better fit than the re-scaled SM prediction, shown as a blue dotted line. The value $m_{\phi/V} =  (2.1\pm 0.1)$~GeV provides the best fit to the data, with a significance of $4.5\sigma$ over the SM.

\begin{figure}[t!]
  \centering
  \includegraphics[width=\columnwidth]{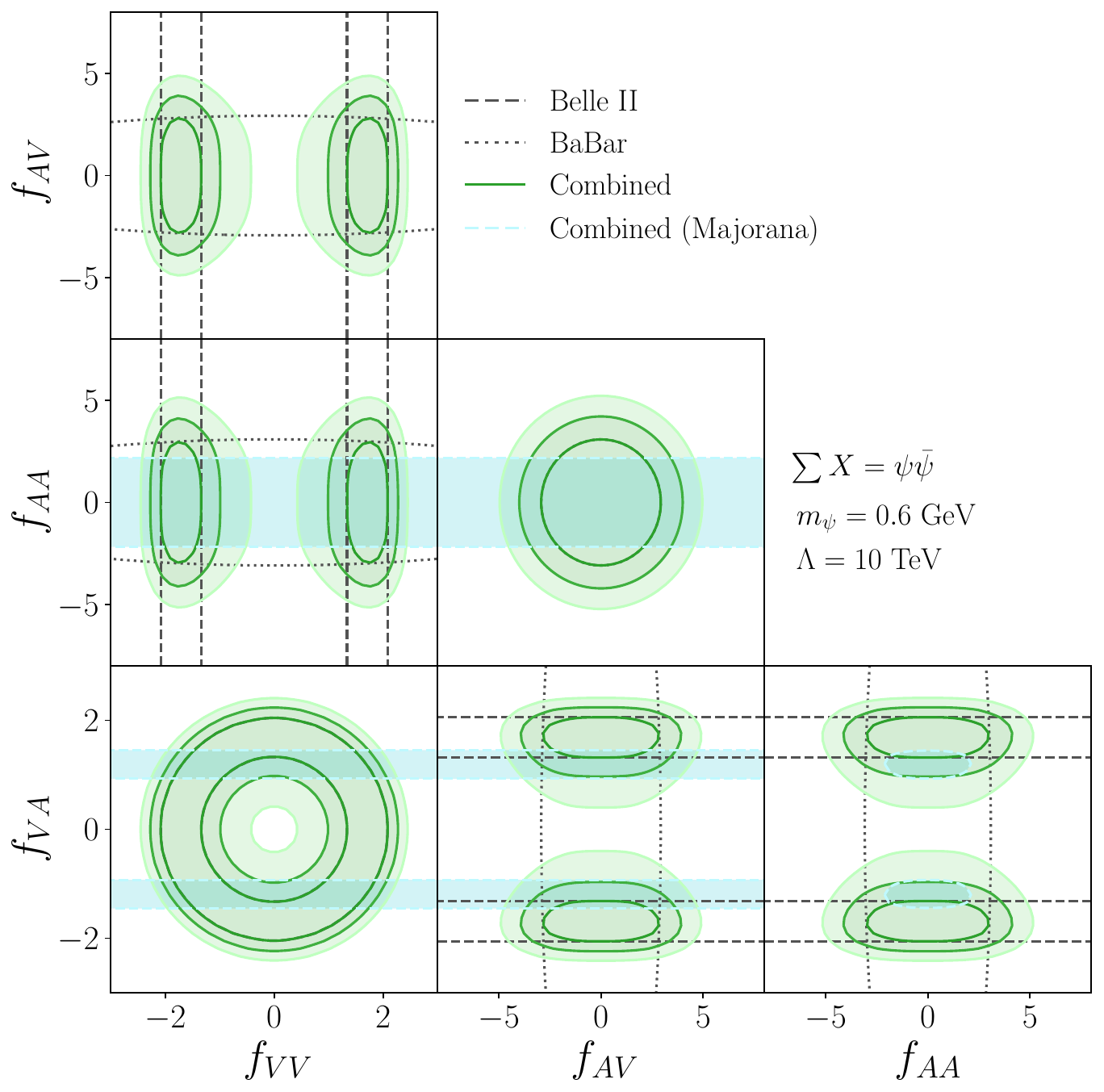}
  \caption{Two-dimensional favoured regions in the parameter space $(f_{VV}, f_{VA}, f_{AV}, f_{AA})$ relevant to the three-body decay $B\to K^{(*)}\psi\bar\psi$, with the mass of $\psi$ fixed to a value close to the best-fit point, $m_\psi = 0.6$~GeV and the heavy NP scale fixed at $\Lambda=10$~TeV. Colour coding is the same as Fig.~\ref{fig:two-body-scalar-coupling-constraints}, with green lines denoting the 1-, 2- and 3-$\sigma$ combined confidence regions instead.
  Additionally, the cyan dashed region corresponds to the 1-$\sigma$ combined confidence region in the case that $\psi$ is a Majorana fermion.}
  \label{fig:three-body-fermion-coupling-constraints}
\end{figure}

In Fig.~\ref{fig:three-body-fits} (above), we show $t_\psi|_{\text{min}}$ for the scenario of the SM plus an invisible fermion pair, $\sum X = \psi\bar{\psi}$. Again, for different values of the fermion mass $m_\psi$, $t_\psi$ is minimised with respect to $(\boldsymbol{\theta}_x,\tau_b)$ and one $f_{XY}$ coupling, with the others set to zero. For masses $m_\psi \lesssim 0.85$~GeV, the vector couplings $f_{VV}$ and $f_{VA}$ provide an improved fit compared to the re-scaled SM, with $m_\psi = 0.60_{-0.14}^{+0.11}$~GeV giving a highest significance of $3.7\sigma$. While still yielding an improvement over the SM-only hypothesis, the scalar and tensor couplings are not as competitive. In Fig.~\ref{fig:three-body-fits} (below), we show $t_X|_{\text{min}}$ for the other three-body decay scenarios; the SM plus pairs of scalars, vectors or spin 3/2 fermions, $\sum X = \phi\bar{\phi}, V\bar{V}, \Psi\bar{\Psi}$. For the scalar pair, both of the couplings $g_{SS}$ (orange, solid) and $g_{VV}$ (orange, dashed) provide a better fit than the re-scaled SM for masses in the ranges $0.05~\text{GeV} \lesssim m_\phi \lesssim 0.76~\text{GeV}$ and $m_\phi \lesssim 0.63~\text{GeV}$, respectively. The scalar masses $m_\phi = 0.52^{+0.11}_{-0.14}$~GeV and $m_\phi = 0.38^{+0.13}_{-0.15}$~GeV provide the best fits at $3.4\sigma$. It is clear from Fig.~\ref{fig:three-body-fits} that the three-body vector (red, solid) and spin $3/2$ (purple, solid) scenarios do not provide a good fit to the Belle~II excess, as expected from the $q^2$ distributions in Fig.~\ref{fig:diff-BRs}.

From Figs.~\ref{fig:two-body-fits} and \ref{fig:three-body-fits}, we see that only the two-body $B\to K\phi/V$ and three-body $B\to K\phi\bar{\phi}/\psi\bar{\psi}$ scenarios provide good fits to the Belle~II excess. For these cases, we now investigate which couplings are favoured. To do so, we turn on two couplings at a time and calculate the profile log-likelihood, $\hat{t}_X = t_{X} - t_{X}|_{\text{min}}$, where $t_X$ is calculated with the couplings at a particular point in the parameter space (minimising over the nuisance parameters) and $t_{X}|_{\text{min}}$ is the likelihood at the global minimum, i.e., minimised with respect to all parameters. However, in both cases, we keep the SM contribution fixed, $\mu = 1$.

Starting with the $B\to K \phi$ mode, we show in Fig.~\ref{fig:two-body-scalar-coupling-constraints} the constraints in the $(g_S, g_P)$ parameter space, with the scalar mass fixed to $m_\phi = 2.1$~GeV. The separate Belle~II and BaBar 1-$\sigma$ regions are shown as grey dotted and dashed lines, respectively. The 1-, 2- and 3-$\sigma$ combined confidence regions are depicted as dark, medium and light orange-shaded contours. Belle~II favours the coupling $|g_S| = (1.6\pm 0.2)\times 10^{-8}$, while BaBar imposes the upper limit $|g_{P}| < 2.5 \times 10^{-8}$ (90\%~CL). In Fig.~\ref{fig:three-body-scalar-coupling-constraints}, we instead show the favoured region in the $(g_{SS}, g_{PS})$ and $(g_{VV}, g_{AV})$ parameter spaces, relevant for $B\to K^{(*)}\phi\bar{\phi}$ and $B_s\to \phi\bar{\phi}$ processes. For $\Lambda = 10$~TeV and the scalar mass fixed to $m_\phi = 0.52$~GeV and $0.38$~GeV for the scalar and vector couplings, respectively, the Belle~II data favours the coupling values $|g_{SS}| = (9.3\pm 1.4)\times 10^{-4}$ and $|g_{VV}| = (3.6\pm 0.5)$, while BaBar and ALEPH put upper bounds on the couplings $g_{PS}$ and $g_{AV}$; ALEPH places a stronger bound on the former, $|g_{PS}|< 2.1\times 10^{-3}$, and BaBar on the latter, $|g_{AV}| < 7.2$, both at 90\%~CL.

\begin{figure}
  \centering
  \includegraphics[width=\columnwidth]{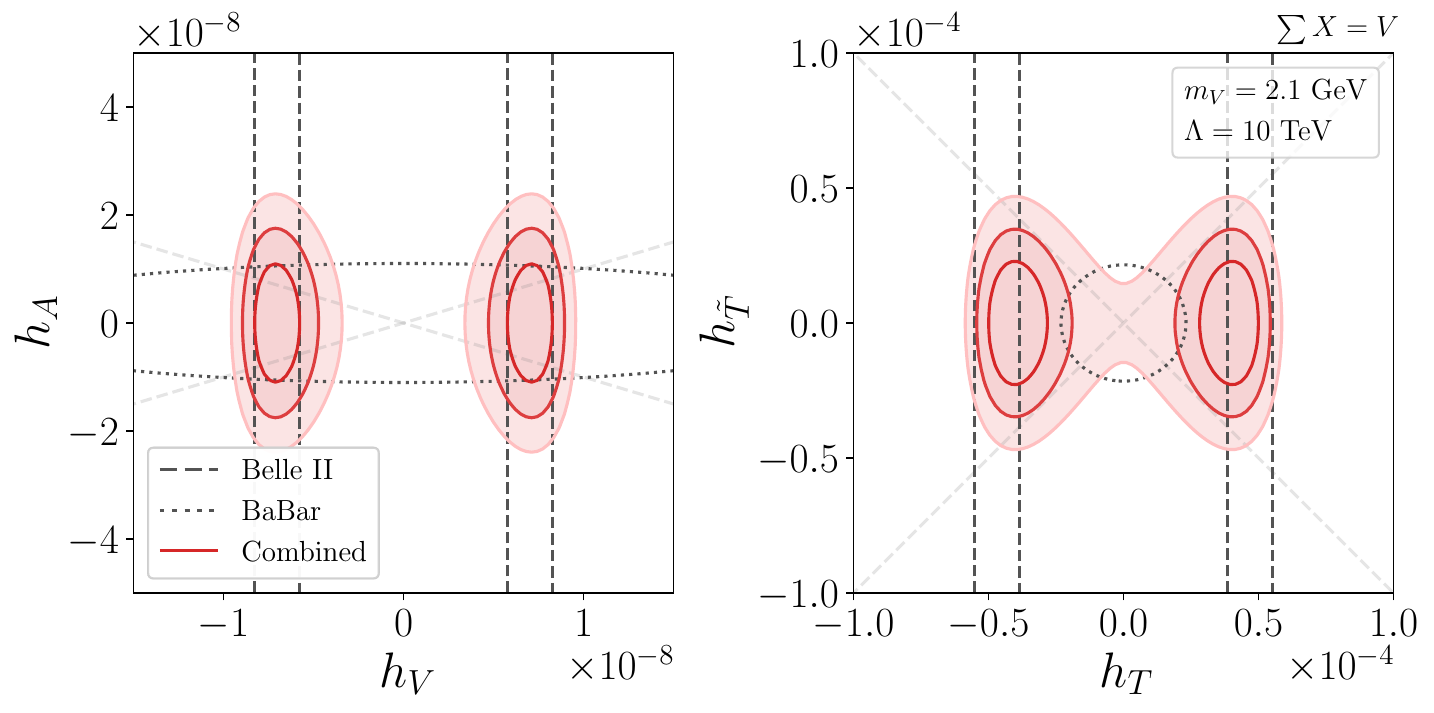}
  \caption{Favoured regions in the parameter space $(h_{V}, h_{A})$ (left) and $(h_{T}, h_{\tilde{T}})$ (right) relevant to the two-body decay $B\to K^{(*)}V$, with the mass of $V$ fixed to a value close to the best-fit point, $m_V = 2.1$~GeV and the heavy NP scale fixed at $\Lambda=10$~TeV. Colour coding is the same as Fig.~\ref{fig:two-body-scalar-coupling-constraints}.
  }
  \label{fig:two-body-vector-coupling-constraints}
\end{figure}

In Fig.~\ref{fig:three-body-fermion-coupling-constraints}, we show the constraints on the parameter space $(f_{VV}, f_{VA}, f_{AV}, f_{AA})$ which enter the decay rates for $B\to K^{(*)}\psi\bar{\psi}$ and $B_s \to \psi\bar{\psi}$. We show planes for each of the six combinations of the couplings, with the others set to zero, for the choices $m_\psi = 0.6$~GeV and $\Lambda = 10$~TeV. Again, the dashed and dotted grey lines show the 1-$\sigma$ regions from Belle~II and BaBar on the invisible Dirac fermion scenario, with the dark, medium and light green regions the 1-, 2- and 3-$\sigma$ combined confidence regions. As seen in the lower left plot of Fig.~\ref{fig:three-body-fermion-coupling-constraints}, the Belle~II data suggests the values $|f_{VV}|^2 + |f_{VA}|^2 = (3.0 \pm 0.8)$ for $f_{AV} = f_{AA} = 0$, while the centre plot shows the BaBar upper bounds $|f_{AV}| < 2.8$ and $|f_{AA}| < 3.0$ for $f_{VV} = f_{VA} = 0$ (90\% CL). The ALEPH bound on $\mathcal{B}(B_s \to \psi\bar{\psi})$ translates to the upper limit $\big|f_{AA}\big| < 13.8$ (90\%~CL), and so BaBar provides more stringent constraints. In Fig.~\ref{fig:three-body-fermion-coupling-constraints}, we also show the 1-$\sigma$ combined region for the invisible Majorana fermion scenario (cyan shaded). These constraints do not depend on the couplings $f_{VV}$ and $f_{AV}$, which vanish.

Finally, in Fig.~\ref{fig:two-body-vector-coupling-constraints}, we show the favoured regions in the $(h_V, h_A)$ and $(h_T, h_{\tilde{T}})$ parameter spaces, which are probed by $B\to K^{(*)} V$, for $m_V = 2.1$~GeV and $\Lambda = 10$~TeV. The colour coding for the Belle~II and BaBar 1-$\sigma$ regions is the same as before, and the dark, medium and light red shaded contours again show the 1-, 2- and 3-$\sigma$ combined confidence regions. Belle~II favours the couplings values $|h_V| = (7.1\pm 0.7)\times 10^{-9}$ and $|h_T| = (4.7\pm 0.5)\times 10^{-5}$, while BaBar enforces the upper bounds $h_A < 7.6 \times 10^{-9}$ (for $h_V$ at the best-fit point) and $|h_{T}|^2 + |h_{\tilde{T}}|^2 < 2.8 \times 10^{-10}$ (90\%~CL). In this case the BaBar data disfavour the $h_T$ value preferred by Belle~II.

{
Several studies have previously considered the Belle II excess in terms of new invisible particles~\cite{Berezhnoy:2023rxx, Felkl:2023ayn, Chen:2023wpb, Dreiner:2023cms, Altmannshofer:2023hkn, He:2024iju}. However, most derive their results  solely from $B\to K^{(*)} E_{\rm miss}$ integrated branching fractions and only one study considered also the decay $B_s \to E_{\rm miss}$~\cite{Felkl:2023ayn}. Only in Ref. \cite{Fridell:2023ssf} was a binned likelihood analysis of $B\to K E_{\rm miss}$ performed within different new physics scenarios relying on both Belle II and BaBar data. 
}
{
We observe that the preferred mass $m_{\phi/V}$  and $m_{\psi}$ obtained from the two- and three-body decay fits respectively are compatible with Refs.~\cite{Altmannshofer:2023hkn,Fridell:2023ssf}. Additionally, our constraint on the $h_V$ and $h_A$ couplings is compatible with the results from Ref.~\cite{Altmannshofer:2023hkn}\footnote{The normalisation of Ref.~\cite{Altmannshofer:2023hkn} differs by a factor of 2, obtaining $g_{V(A)}^{(4)}=2h_{V(A)}$.}.}
On the other hand, our estimated significances of NP scenarios over the SM differ somewhat from those given in Ref~\cite{Fridell:2023ssf}. In particular, we find that two-body decay kinematics seems to give a better fit to Belle II data than three-body decay spectra.

%%%%%%%%%%%%%%%%%%%%%%%%%%%%%%%%%%%%%%%%%%
%
\section{Conclusions}
%
%%%%%%%%%%%%%%%%%%%%%%%%%%%%%%%%%%%%%%%%%%

%We find that the signal of Belle II is probably due to B->K (D-> K0 K0) :-P 

\label{sec:conclusions}In the present work, we have explored the interesting possibility that the signal of $B\to K E_{\rm miss}$ recently observed by Belle II arises not only from the SM neutrinos but from new undetected particles in the final state. By taking into account available experimental data on the total event yields, as well as on the differential $B\to K^{(*)} E_{\rm miss}$ distributions presented in Belle II and BaBar analyses, we were able to construct the likelihood for different NP scenarios using experimental information from both processes. {In addition, we considered constraints on the branching ratio $B_s \to E_{\rm miss}$ obtained from the ALEPH data.}

We considered possible decay channels involving one new invisible scalar or vector state as well as decays to two invisible states, either scalars, spin-$1/2$ fermions, vectors or even spin-$3/2$ states. 

The two-body $B\to K^{(*)} X$ topology with an invisible particle mass of $m_X=2.1$~GeV can accommodate all data much better than the (re-scaled) SM. In particular, the Belle II data prefer these scenarios by ($3.5\sigma$) $4.5\sigma$, respectively. This conclusion seems to differ somewhat from the results of Ref.~\cite{Fridell:2023ssf}. 
 
In the case of two invisible new particles in the final state, i.e. $B\to K^{(*)} X \bar X$, we find that two scalars or two spin-$1/2$ fermions can also accommodate all data with the most favoured masses in the range $m_X=0.4-0.6$~GeV, depending on the $X$ spin and couplings considered. Conversely, scenarios with pairs of vectors or spin $3/2$ fermions in the final state cannot improve the SM fit of Belle II and are thus disfavoured irrespective of their mass or couplings. 

In the parity basis of effective operators, we find that scenarios with spin-$1/2$ fermions coupled to the vector quark current are preferred over those coupling to scalar or tensor quark currents. On the other hand, scenarios with pairs of scalar particles coupled to vector or scalar quark currents are equally likely. 

The results of the BaBar search for $B\to K^{*} E_{\rm miss}$ are consistent with most scenarios preferred by Belle II data. In the parameter space of possible couplings, they yield predominantly orthogonal constraints. One notable exception is a single massive vector field coupled to the tensor (dipole) quark current. The Belle II and BaBar results cannot be accommodated simultaneously in this scenario. {Finally, the ALEPH data on $B_s \to E_{\rm miss}$ currently provide the strongest constraint on a pair of scalars coupled to a pseudoscalar quark bilinear operator ($\bar s \gamma_5 b$).} 

The impact of $B\to K^{*} E_{\rm miss}$ (and $B_s \to  E_{\rm miss}$) constraints is somewhat more interesting in the chiral basis of quark operators (see also Appendix~\ref{app:basis}), which is often preferred for UV model building respecting the SM chiral gauge symmetry. Interestingly, we find that purely chiral couplings still lie within the combined $68\%$ CL region of both Belle II and BaBar (ALEPH) results. However, a future improvement on $B\to K^{*} E_{\rm miss}$ by Belle II could soon put most chiral scenarios under pressure. 

We also note that coupling ranges implied by our fit are mostly consistent with NP scales even above our benchmark $\Lambda=10$~TeV. 
Consequently, new degrees of freedom associated with higher dimensional operator terms in Eqs.~\eqref{eq:L_V},~\eqref{eq:L_S}, and~\eqref{eq:L_T} could lie above the LHC reach. Notable exceptions are pairs of invisible scalars or fermions coupled to the quark vector current ($\bar s \gamma_\mu b$), where the fit to Belle II signal points to NP below few TeV.  

While we only considered $b \to s$ transitions in this work, in more complete NP scenarios, other phenomenology could become relevant. In particular, NP flavour constructions (see e.g.~\cite{Badin:2010uh, Bauer:2021mvw, Li:2023sjf, He:2024iju}) would relate these decays to $K\to \pi E_{\rm miss}$, $D\to \pi E_{\rm miss}$ and monojet/monotop production at high $p_T$, leading to possibly strong constraints from NA62~\cite{NA62:2021zjw}, KOTO~\cite{KOTO:2020prk}, BESSIII~\cite{BESIII:2021slf} and LHC~\cite{CMS:2014ofj, ATLAS:2024vqf} experiments. 
The results of our Belle II likelihood fit however indicate that for the preferred masses of invisible states, they could be kinematically forbidden from being produced in rare kaon and possibly even $D$ meson decays.
Finally, in UV complete models, constraints from $B_s$ meson oscillations\footnote{In scenarios with a single scalar or vector, these arise at the tree level, but have been found in Ref.~\cite{Bauer:2020jbp} to be negligible, a result which we have also verified explicitly.} as well as EW precision data could be relevant. We leave these considerations for future work.

%%%%%%%%%%%%%%%%%%%%%%%%%%%%%%%%%%%%%%%%%%

\begin{acknowledgments}
We thank Luka \v Santelj and Peter Kri\v zan for multiple illuminating discussions regarding the details of the Belle~II experimental analysis and in particular Luka \v Santelj for providing the relevant smearing and efficiency maps of the Belle~II ITA tagging analysis. We also thank Damir Be\v cirevi\' c and Olcyr Sumensari for providing useful information on the relevant $B\to K^{(*)}$ form factors. PDB, SF and JFK acknowledge financial support from the Slovenian Research Agency (research core funding No. P1-0035, J1-3013 and N1-0321). MN acknowledges the financial support by the Spanish Government (Agencia Estatal de Investigaci\'on MCIN/AEI/10.13039/501100011033)  and the European Union NextGenerationEU/PRTR through the “Juan de la Cierva” program (Grant No. JDC2022-048787-I)
and Grant No. PID2020-114473GB-I00. MN also acknowledges the support of the Generalitat Valenciana through Grant No. PROMETEO/2021/071. This study has been partially carried out within the INFN project (Iniziativa Specifica) QFT-HEP.
\end{acknowledgments}
\vspace{1em}
\appendix
\vspace{-1em}
%%%%%%%%%%%%%%%%%%%%%%%%%%%%%%%%%%%%%%%%%%
%
\section{Chiral Operator Basis}
\label{app:basis}
%
%%%%%%%%%%%%%%%%%%%%%%%%%%%%%%%%%%%%%%%%%%

To convert from the parity basis (Eqs.~\eqref{eq:L_V},~\eqref{eq:L_S}, and~\eqref{eq:L_T}) to the chiral basis (defined in Ref.~\cite{Kamenik:2011vy}) for the operators involving the fields $X \in \{\phi, \psi, V_\mu, \Psi_\mu\}$, it is convenient to define the following matrices,
\begin{align}
P = \frac{1}{2}
\begin{pmatrix}
1 & 1 \\
-1 & 1
\end{pmatrix}\,,~~ Q = \frac{1}{4}\begin{pmatrix}
1 & 1 & 1 & 1 \\
-1 & 1 & -1 & 1 \\
-1 & -1 & 1 & 1 \\
1 & -1 & -1 & 1 \\
\end{pmatrix}\,.
\end{align}
For the operators containing the dark scalar field $\phi$ in Eqs.~\eqref{eq:L_V} to \eqref{eq:L_T}, we then have the matching relations,
\begin{gather}
\begin{pmatrix}
g_{S(S)} \\
g_{P(S)} \\
\end{pmatrix} = 
P\frac{v}{\sqrt{2}\Lambda}
\begin{pmatrix}
C_{d\phi(\phi)}^{S,L} \\
C_{d\phi(\phi)}^{S,R} \\
\end{pmatrix}\,, \nonumber\\
\begin{pmatrix}
g_{VV} \\
g_{AV} \\
\end{pmatrix} = 
P \begin{pmatrix}
C_{d\phi\phi}^{V,L} \\
C_{d\phi\phi}^{V,R} \\
\end{pmatrix}\,.
\end{gather}
For the four-fermion operators involving the invisible fermion field $\psi$, we have
\begin{gather}
\begin{pmatrix}
f_{VV} \\
f_{VA} \\
f_{AV} \\
f_{AA} \\
\end{pmatrix} = 
Q \begin{pmatrix}
C_{d\psi}^{V,LL} \\
C_{d\psi}^{V,LR}  \\
C_{d\psi}^{V,RL}  \\
C_{d\psi}^{V,RR}  \\
\end{pmatrix}\,, ~\begin{pmatrix}
f_{SS} \\
f_{SP} \\
f_{PS} \\
f_{PP} \\
\end{pmatrix} = 
Q\frac{v}{\sqrt{2}\Lambda}\begin{pmatrix}
C_{d\psi}^{S,LL} \\
C_{d\psi}^{S,LR} \\
C_{d\psi}^{S,RL} \\
C_{d\psi}^{S,RR} \\
\end{pmatrix}\,, \nonumber \\
\begin{pmatrix}
f_{TT} \\
f_{\tilde{T}T} \\
\end{pmatrix} = 
P\frac{v}{\sqrt{2}\Lambda}
 \begin{pmatrix}
C_{d\psi}^{T,LL} \\
C_{d\psi}^{T,RR} \\
\end{pmatrix}\,.\label{eq:chiraltoparity}
\end{gather}
The same relations hold for the effective couplings of the spin 3/2 field $\Psi_\mu$, with the replacements $f_{XY}\to F_{XY}$ and $\psi \to \Psi$. Additionally, the couplings $F_{TS}$, $F_{TP}$, $F_{\tilde{T}S}$ and $F_{\tilde{T}P}$ are related to chiral basis coefficients in a similar manner as the scalar and pseudoscalar couplings. Finally, the operators coupling the vector field $V_\mu$ to the relevant quark couplings are rotated to the chiral basis as
\begin{gather}
\begin{pmatrix}
h_{V} \\
h_{A} \\
\end{pmatrix} = 
P
\begin{pmatrix}
C_{dV}^{V,L} \\
C_{dV}^{V,R} \\
\end{pmatrix}\,, \quad \begin{pmatrix}
h_{T} \\
h_{\tilde{T}} \\
\end{pmatrix} = 
P\frac{v}{\sqrt{2}\Lambda}
\begin{pmatrix}
C_{dV}^{T,L} \\
C_{dV}^{T,R} \\
\end{pmatrix}\,, \nonumber \\
\begin{pmatrix}
h_{S} \\
h_{P} \\
\end{pmatrix} = 
P \frac{v}{\sqrt{2}\Lambda} \begin{pmatrix}
C_{dVV}^{S,L} \\
C_{dVV}^{S,R} \\
\end{pmatrix}\,.
\end{gather}

In Figs.~\ref{fig:two-body-scalar-coupling-constraints}, \ref{fig:three-body-scalar-coupling-constraints} and \ref{fig:two-body-vector-coupling-constraints}, the constraints on the chiral-basis couplings can be inferred from the light grey dashed lines, which show, for example, $g_S = g_P$ and $g_S = - g_P$ (which equivalently depict $C_{d\phi}^{S,L} = 0$ and $C_{d\phi}^{S,R} = 0$, respectively). 

\begin{figure}[t!]
  \centering
  \includegraphics[width=\columnwidth]{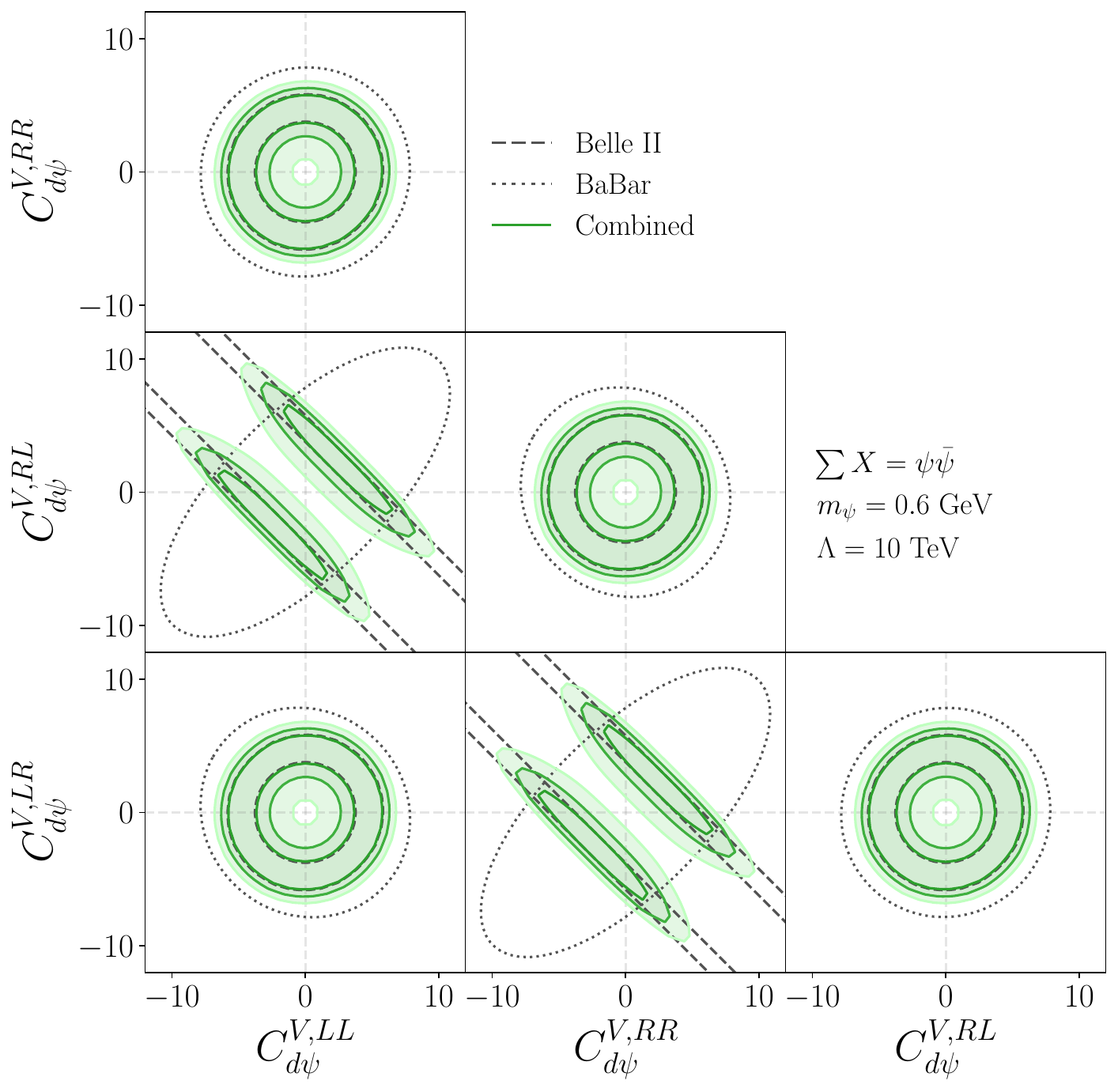}
  \caption{Two-dimensional favoured regions in the parameter space $(C_{d\psi}^{LL}, C_{d\psi}^{LR}, C_{d\psi}^{RL}, C_{d\psi}^{RR})$ relevant to the three-body decay $B\to K^{(*)}\psi\bar\psi$, with the mass of $\psi$ fixed to a value close to the best-fit point, $m_\psi = 0.6$~GeV and the heavy NP scale fixed at $\Lambda=10$~TeV. Colour coding is the same as Fig.~\ref{fig:two-body-scalar-coupling-constraints}, with green lines denoting the 1-, 2- and 3-$\sigma$ combined confidence regions.}
  \label{fig:three-body-fermion-coupling-constraints-chiral}
\end{figure}

The transformation to the chiral basis for the constraints on the fermion couplings in Fig.~\ref{fig:three-body-fermion-coupling-constraints} is slightly less trivial than a rotation because two of couplings $(f_{VV}, f_{VA}, f_{AV}, f_{AA})$ are taken to be zero in each of the six plots. We instead perform an additional fit for the chiral couplings $(C_{d\psi}^{LL}, C_{d\psi}^{LR}, C_{d\psi}^{RL}, C_{d\psi}^{RR})$, with the favoured regions shown in Fig.~\ref{fig:three-body-fermion-coupling-constraints-chiral}.
The dashed and dotted grey lines again show the 1-$\sigma$ regions from Belle~II and BaBar, while the dark, medium and light green contours show the 1-, 2- and 3-$\sigma$ combined confidence regions.

%

%%%%%%%%%%%%%%%%%%%%%%%%%%%%%%%%%%%%%%%%%%
%
\section{Analytic Expressions of Decay Rates}
\label{app:decays}
%
%%%%%%%%%%%%%%%%%%%%%%%%%%%%%%%%%%%%%%%%%%

In this appendix, we give explicit expressions for the relevant two-body and three-body $B$ decays to $K^{(*)}$ and the invisible final states $X\in \{\phi, \psi, V, \Psi\}$, first explored in Ref.~\cite{Kamenik:2011vy}. We also give the decay rates for invisible two-body $B_s$ decays. The results make use of the form factors and decay constants defined in Appendix~\ref{app:ffs}.

%%%%%%%%%%%%%%%%%%%%%%%%%%%%%%%%%%%%%%%%%%
\subsection{$B\to K$ Two-Body Decays}
%%%%%%%%%%%%%%%%%%%%%%%%%%%%%%%%%%%%%%%%%%

The two-body decay rates for $B\to K^{(*)}\phi$ are
\begin{align}
\Gamma(B\to K \phi) &= \frac{|g_S|^2}{8\pi}\frac{\left|\vec{p}_{K}\right| m_B^2\delta_K^2}{(m_b - m_s)^2} f_{0}^2(m_\phi^2)\,, \label{eq:BKphi} \\
\Gamma(B\to K^* \phi) &= \frac{|g_P|^2}{2\pi}\frac{\left|\vec{p}_{K^{*}}\right|^3}{(m_b + m_s)^2} A_{0}^2(m_\phi^2)\,, \label{eq:BKsphi}
\end{align}
where $\delta_{K^{(*)}} \equiv (1 - m_{K^{(*)}}^2/m_B^2)$, $g_S$ ($g_P$) is the scalar (pseudoscalar) coupling of the scalar field $\phi$ to the flavour-changing quark current $\bar{s}b$ ($\bar{s}\gamma_5 b$), and the three-momentum of $K^{(*)}$ is
\begin{align}
\label{eq:K-3momentum}
\left|\vec{p}_{K^{(*)}}\right| = \frac{\lambda^{1/2}(m_B^2,m_{\phi}^2,m_{K^{(*)}}^2)}{2m_B}\,,
\end{align}
with $\lambda(x,y,z) = (x - y - z)^2 - 4 y z$. Note that the above rates also apply to the operators $(\bar{s}\gamma_\mu b)\partial^\mu \phi/\Lambda$ and $(\bar{s}\gamma_\mu \gamma_5 b)\partial^\mu \phi/\Lambda$ if one makes the replacement $g_{S(P)} \to i m_b g_{V(A)}/\Lambda$.

\begin{widetext}

The two-body decay rates for $B\to K^{(*)}V$ are instead given by
\begin{align}
\Gamma(B\to K V) &= \frac{|\vec{p}_K|}{2\pi}\Bigg[|h_{V}|^2\frac{|\vec{p}_K|^2}{m_V^2} f_+^2(m_V^2) +  4|h_{T}|^2 \frac{m_V^2}{\Lambda^2}\frac{|\vec{p}_K|^2}{(m_B+m_K)^2}f_T^{2}(m_V^2)  
\nonumber \\
& \hspace{4em} + 4\,\Re[h_V h_T^*]\frac{|\vec{p}_K|^2}{(m_B+m_K) \Lambda}f_+(m_V^2)f_T(m_V^2)\Bigg]\,, \\
\Gamma(B\to K^* V) &= \frac{|\vec{p}_{K^*}|}{2\pi} \Bigg[2|h_{V}|^2 \frac{|\vec{p}_{K^*}|^2}{(m_B + m_{K^*})^2} V^2(m_V^2) + |h_{A}|^2\left(\frac{(m_B + m_{K^*})^2}{2m_B^2} A_{1}^2(m_V^2) + \frac{16 m_{K^*}^2}{m_V^2}A_{12}^2(m_V^2)\right) \nonumber \\
& \hspace{4em} + 8|h_{T}|^2\frac{|p_{K^*}|^2}{\Lambda^2} T_{1}^{2}(m_V^2) + 2|h_{\tilde{T}}|^2\frac{m_V^2}{\Lambda^2} \bigg(\frac{m_B^2\delta_{K^*}^2}{m_V^2} T_{2}^2(m_V^2) + \frac{8 m_{K^*}^2}{(m_B + m_{K^*})^2} T_{23}^2(m_V^2)\bigg) \nonumber\\
&\hspace{4em} - 8\,\Re[h_V h_T^*]\frac{|\vec{p}_{K^*}|^2}{(m_B + m_{K^*}) \Lambda}V(m_V^2)T_1(m_V^2) \nonumber\\ 
&\hspace{4em} - 2\,\Re[h_A h_{\tilde{T}}^*]\frac{m_B + m_{K^*}}{\Lambda}\bigg(\delta_{K^*} A_1(m_V^2) T_2(m_V^2) + \frac{16m_{K^*}^2}{(m_B  + m_{K^*})^2}A_{12}(m_V^2)T_{23}(m_V^2)\bigg) \Bigg]\,,
\end{align}
where $h_V$ ($h_A$) is the (axial) vector coupling of the vector field $V_\mu$ to the flavour-changing quark current $\bar{s}\gamma_\mu b$ ($\bar{s}\gamma_\mu \gamma_5 b$) and $h_T$ ($h_{\tilde{T}}$) is the magnetic (electric) dipole-like coupling of the field strength tensor $V^{\mu\nu}$ to the flavour-changing quark current $\bar{s}\sigma_{\mu\nu}b$ ($\bar{s}\sigma_{\mu\nu}\gamma_5 b$). The three-momentum of $K^{(*)}$ is given by Eq.~\eqref{eq:K-3momentum} with the replacement $m_\phi \to m_V$. 

%%%%%%%%%%%%%%%%%%%%%%%%%%%%%%%%%%%%%%%%%%
\subsection{$B\to K$ Three-Body Decays}
%%%%%%%%%%%%%%%%%%%%%%%%%%%%%%%%%%%%%%%%%%

Firstly, the three-body differential decay rates in $q^2$ for $B\to K^{(*)}\phi\bar{\phi}$ are given by
\begin{align}
\label{eq:BK_phiphi}
\frac{d\Gamma(B\to K \phi\bar{\phi})}{dq^2} &= \frac{\beta_{\phi}}{96\pi^3}\frac{|\vec{p}_K|}{\Lambda^2}\Bigg[\frac{3}{4}|g_{SS}|^2\frac{m_B^2\delta_{K}^2}{(m_b - m_s)^2} f_{0}^2(q^2) + |g_{VV}|^2\frac{|\vec{p}_K|^2}{\Lambda^2} \beta_{\phi}^2  f_{+}^2(q^2)\Bigg]\,,  \\
\frac{d\Gamma(B\to K^* \phi\bar{\phi})}{dq^2} &= \frac{\beta_{\phi}}{96\pi^3}\frac{|\vec{p}_{K^*}|}{\Lambda^2}\Bigg[3|g_{PS}|^2\frac{|\vec{p}_{K^*}|^2}{(m_b + m_s)^2}A_{0}^2(q^2) + 2|g_{VV}|^2\frac{q^2}{\Lambda^2} \beta_{\phi}^2 \frac{|\vec{p}_{K^*}|^2}{(m_B + m_{K^*})^2} V^2(q^2) \nonumber\\
& \hspace{7.5em} + |g_{AV}|^2  \frac{q^2}{\Lambda^2} \beta_{\phi}^2 \left(\frac{(m_B + m_{K^*})^2}{2m_B^2} A_{1}^2(q^2) + \frac{16m_{K^*}^2}{q^2}A_{12}^2(q^2)\right)\Bigg] \,,
\end{align}
where $\beta_X = \sqrt{1 - 4m_X^2/q^2}$ and the three-momentum of $K^{(*)}$ is given by Eq.~\eqref{eq:K-3momentum} with the replacement $m_\phi^2 \to q^2$. In the scenario of a real scalar field, where $\phi = \bar{\phi}$, the contributions from $g_{VV}$ and $g_{AV}$ vanish, and the remaining contributions from $g_{SS}$ and $g_{SP}$ are a factor of $2$ larger at the amplitude level; hence, the replacements $g_{SS} \to 2 g_{SS}$ and $g_{PS} \to 2 g_{PS}$ can be made in Eq.~\eqref{eq:BK_phiphi}. The rate must also be multiplied by factor of $1/2$ to account for identical outgoing states.

The differential decay rate for $B\to K\psi\bar{\psi}$ is
\begin{align}
\label{eq:BK_psipsi}
\frac{d\Gamma(B\to K \psi\bar{\psi})}{dq^2} & =  \frac{\beta_{\psi}}{24\pi^3}\frac{|\vec{p}_K|q^2}{\Lambda^4} \nonumber \\
&\hspace{-6.5em}\times \Bigg[\bigg(\Big(|f_{VV}|^2\beta_\psi^{\prime 2} + |f_{VA}|^2\beta_{\psi}^2\Big)f_+(q^2) + 12\,\Re\big[f_{VV}f_{TT}^*\big] \frac{m_\psi}{m_B + m_K}f_{T}(q^2)\bigg)\frac{|\vec{p}_K|^2}{q^2} f_{+}(q^2) \nonumber \\
& \hspace{-5em} + \frac{3}{8}\bigg(4|f_{VA}|^2\frac{m_\psi^2(m_b - m_s)^2}{q^4} + |f_{SS}|^2\beta_{\psi}^2 + |f_{SP}|^2 + 4\,\Re\big[f_{VA}f_{SP}^*\big]\frac{m_\psi (m_b - m_s)}{q^2}\bigg) \frac{m_B^2\delta_K^2}{(m_b - m_s)^2} f_{0}^2(q^2) \nonumber \\
& \hspace{-5em} + 2\Big(|f_{TT}|^2 \beta_\psi^{\prime\prime 2} + |f_{\tilde{T}T}|^2 \beta_\psi^2\Big) \frac{|\vec{p}_K|^2}{(m_B + m_K)^2}f_{T}^2(q^2) \Bigg]\,, 
\end{align}
where we have defined $\beta_\psi^{\prime 2} = (3-\beta_\psi^2)/2 = (1 + 2m_\psi^2/q^2)$ and $\beta_\psi^{\prime\prime 2} = (3-2\beta_\psi^2) = (1 + 8m_\psi^2/q^2)$. Likewise, the decay rate for $B\to K^{*}\psi\bar{\psi}$ is
\begin{align}
\label{eq:BKs_psipsi}
\frac{d\Gamma(B\to K^{*} \psi\bar{\psi})}{dq^2} &= \frac{\beta_{\psi}}{24\pi^3}\frac{|\vec{p}_{K^*}|q^2}{\Lambda^4} \nonumber \\
& \hspace{-6.5em}\times \Bigg[2\Big(|f_{VV}|^2\beta_\psi^{\prime 2} + |f_{VA}|^2\beta_\psi^2\Big) \frac{|\vec{p}_{K^*}|^2}{(m_B + m_{K^*})^2} V^{2}(q^2) \nonumber \\
& \hspace{-4.5em} + \Big(|f_{AV}|^2\beta_\psi^{\prime 2} + |f_{AA}|^2\beta_\psi^{2}\Big) \left(\frac{(m_B + m_{K^*})^2}{2m_B^2} A_{1}^2(q^2) + \frac{16m_{K^*}^2}{q^2}A_{12}^2(q^2)\right) \nonumber \\
& \hspace{-4.5em} + \frac{3}{2}\bigg(4|f_{AA}|^2\frac{m_\psi^2(m_b + m_s)^2}{q^4} + |f_{PS}|^2 \beta_{\psi}^2 + |f_{PP}|^2 - 4\,\Re\big[f_{AA}f_{PP}^*\big] \frac{m_\psi(m_b+m_s)}{q^2}\bigg)\frac{|\vec{p}_{K^*}|^2}{(m_b + m_s)^2} A_0^{2}(q^2) \nonumber \\
& \hspace{-4.5em} + 4\bigg(\Big(|f_{TT}|^2\beta_\psi^{\prime\prime 2} + |f_{\tilde{T}T}|^2\beta_\psi^{2}\Big)T_{1}(q^2) - 6\,\Re\big[f_{VV}f_{TT}^*\big]\frac{m_\psi}{m_B + m_{K^*}}V(q^2)\bigg)\frac{|\vec{p}_{K^*}|^2}{q^2}T_{1}(q^2)\nonumber \\
& \hspace{-4.5em} + \Big(|f_{TT}|^2\beta_\psi^{2} + |f_{\tilde{T}T}|^2\beta_\psi^{\prime\prime 2}\Big)\bigg(\frac{m_B^2\delta_{K^*}^2}{q^2} T_2^2(q^2) + \frac{8m_{K^*}^2}{(m_B + m_{K^*})^2}T_{23}^2(q^2)\bigg) \nonumber \\
& \hspace{-4.5em} -6\,\Re\big[f_{AV}f_{\tilde{T}T}^*\big] \frac{m_\psi (m_B + m_{K^*})}{q^2}\left(\delta_{K^*} A_1(q^2) T_2(q^2) + \frac{16m_{K^*}^2}{(m_B + m_{K^*})^2}A_{12}(q^2) T_{23}(q^2)\right)\Bigg]\,.
\end{align}
The differential decay rates in Eq.~\eqref{eq:BK_psipsi} and~\eqref{eq:BKs_psipsi} are valid for a pair of invisible Dirac fermions in the final state. However, the result for Majorana fermions 
($\psi = \psi^c$) can be obtained from Eq.~\eqref{eq:BK_psipsi} and~\eqref{eq:BKs_psipsi} with the replacements $f_{VV}, f_{AV}, f_{TT}, f_{\tilde{T}T}\to 0$ and $f_{XY} \to 2 f_{XY}$ for the remaining couplings. Additionally, a factor of $1/2$ is required to account for identical outgoing states.

The differential decay rate for the process $B\to K V\bar{V}$ is
\begin{align}
\frac{d\Gamma(B\to K V\bar{V})}{dq^2} &= \frac{\beta_V|h_S|^2}{512\pi^3}\frac{|\vec{p}_K|q^4}{m_V^2 \Lambda^2}\mathcal{J}_V\frac{m_B^2\delta_K^2}{(m_b - m_s)^2}f_0^2(q^2)\,,
\end{align}
where $\mathcal{J}_V = (3-2\beta_V^2 + 3\beta_V^4)/4$. Again, for $V = \bar{V}$, the replacement $h_S \to 2 h_S$ must be made and the rate multiplied by $1/2$ for the reduction in phase space.

Finally, the differential decay rate for the process $B\to K\Psi\bar{\Psi}$ is
\begin{align}
\label{eq:BK_PsiPsi}
\frac{d\Gamma(B\to K \Psi\bar{\Psi})}{dq^2} &= \frac{\beta_\Psi}{216\pi^3}\frac{|\vec{p}_{K}|q^6}{m_\Psi^4\Lambda^4} \nonumber \\
&\hspace{-6.5em}\times \Bigg[\bigg(\Big(|F_{VV}|^2\mathcal{J}_{VV} + |F_{VA}|^2\beta_{\Psi}^2\mathcal{J}_{VA}\Big)f_+(q^2) \nonumber \\
& \hspace{-4em} + 2\,\Re\Big[F_{VV}\Big(6F_{TT}^* \mathcal{J}_{VV,TT} + 2F_{TS}^*\beta_\Psi^4 + F_{\tilde{T}P}^*\mathcal{J}_{VV,\tilde{T}P}\Big)\Big]\frac{m_\Psi}{m_B + m_K}f_T(q^2)\bigg)\frac{|\vec{p}_K|^2}{q^2} f_{+}(q^2) \nonumber \\
& \hspace{-5em} + \frac{3}{8}\bigg(|F_{VA}|^2(1-\beta_\Psi^2)\mathcal{J}_{\Psi}\frac{(m_b - m_s)^2}{q^2} \nonumber \\
& \hspace{-2.5em} + |F_{SS}|^2\beta_{\Psi}^2\mathcal{J}_{\Psi}' + |F_{SP}|^2\mathcal{J}_{\Psi} + 4\,\Re\big[F_{VA}F_{SP}^*\big]\mathcal{J}_{\Psi}\frac{m_\Psi (m_b - m_s)}{q^2}\bigg) \frac{m_B^2\delta_K^2}{(m_b - m_s)^2} f_{0}^2(q^2) \nonumber \\
& \hspace{-5em} + 2\bigg(|F_{TT}|^2 \mathcal{J}_{TT} + \frac{1}{4}|F_{TS}|^2\beta_\Psi^4\mathcal{J}_{TS} + \frac{1}{4}|F_{TP}|^2\beta_\Psi^2\mathcal{J}_{TP} \nonumber \\
& \hspace{-2.5em} + |F_{\tilde{T}T}|^2 \beta_\Psi^2\mathcal{J}_{\Psi} + \frac{3}{16}|F_{\tilde{T}S}|^2\beta_\Psi^2(1 - \beta_\Psi^2)\mathcal{J}_{\tilde{T}S} + \frac{5}{16}|F_{\tilde{T}P}|^2\beta_{\Psi}^{\prime 2}(1 - \beta_\Psi^2) \nonumber \\
& \hspace{-2.5em} + \Re\bigg[F_{TT}\bigg(F_{TS}^*\beta_\Psi^4 \mathcal{J}_{TT,TS} + \frac{5}{4}F_{\tilde{T}P}^*\beta_{\Psi}^{\prime 2}(1 - \beta_\Psi^2)\bigg) - \frac{1}{2}F_{TP} F_{\tilde{T}S}^*\beta_{\Psi}^2(1-\beta_\Psi^2) \nonumber\\
& \hspace{0em}+ F_{\tilde{T}T}\bigg(F_{TP}^*\beta_\Psi^2 \mathcal{J}_{\tilde{T}T,TP} - \frac{1}{4}F_{\tilde{T}S}^*\beta_{\Psi}^2(1 - \beta_\Psi^2)\mathcal{J}_{\tilde{T}T,\tilde{T}S}\bigg)\bigg]\bigg) \frac{|\vec{p}_K|^2}{(m_B + m_K)^2}f_{T}^2(q^2) \Bigg]\,, 
\end{align}
\end{widetext}
where we have defined for convenience the factors $\mathcal{J}_\Psi = (9 - 6 \beta_\Psi^2 + 5 \beta_\Psi^4)/8$, $\mathcal{J}_\Psi' = (5 - 6 \beta_\Psi^2 + 9\beta_\Psi^4)/8$ and,
\begin{align}
\mathcal{J}_{VV} &= (15 - 15\beta_\Psi^2 + 25\beta_\Psi^4 - 9\beta_\Psi^6)/16 \,, \nonumber \\
\mathcal{J}_{VA} &= (5-2\beta_\Psi^2 + 5\beta_\Psi^4)/8 \,, \nonumber \\
\mathcal{J}_{TT} & = (15 - 20\beta_\Psi^2 + 23 \beta_\Psi^4 - 10\beta_\Psi^6)/8 \,, \nonumber \\
\mathcal{J}_{TS} & = \mathcal{J}_{TT,TS} = (9 - 5\beta_\Psi^2)/4 \,, \nonumber \\
\mathcal{J}_{TP} & = (7 - 3\beta_\Psi^2)/4 \,, ~~ \mathcal{J}_{\tilde{T}S} = (11 - 5\beta_\Psi^2)/6 \,, \nonumber \\
\mathcal{J}_{VV,TT} & = (15 - 10\beta_\Psi^2 + 19\beta_\Psi^4)/24 \,, \nonumber \\
\mathcal{J}_{VV,\tilde{T}P} & = (15 - 10\beta_\Psi^2 + 3\beta_\Psi^4)/8 \,, \nonumber \\
\mathcal{J}_{\tilde{T}T,TP} &= (3 + \beta_\Psi^2)/4 \,, \nonumber \\
\mathcal{J}_{\tilde{T}T,\tilde{T}S} &= (-3 + 5\beta_\Psi^2)/2\,.
\end{align}
For $\Psi = \Psi^c$, we get the decay rate from Eq.~\eqref{eq:BK_PsiPsi} with $F_{VV}, F_{AV}, F_{TT}, F_{\tilde{T}T}\to 0$ and $F_{XY} \to 2 F_{XY}$ for the remaining couplings. Eq.~\eqref{eq:BK_PsiPsi} must also be multiplied by $1/2$.

%%%%%%%%%%%%%%%%%%%%%%%%%%%%%%%%%%%%%%%%%%
\subsection{Invisible $B_s$ Decays}
\label{subsec:inv_Bs}
%%%%%%%%%%%%%%%%%%%%%%%%%%%%%%%%%%%%%%%%%%

In the presence of light hidden states $X\in\{\phi,\psi,V,\Psi\}$, invisible $B_s$ decays, $B_s \to \sum X$, can also occur via the same operators that trigger $B \to K^{*} \sum X$. Here, we give expressions for the induced decay rates, which are used to recast the upper bound on $\mathcal{B}(B_s \to \text{inv})$ from ALEPH onto the effective coupling parameter space. In the following, we make use of the $B_s$ decay matrix elements defined in Appendix~\ref{Bsffs}.

\begin{widetext}
The decay rate for a pair of outgoing invisible scalars is
\begin{align}
\Gamma(B_s \to \phi\bar{\phi}) = \frac{\beta_\phi}{16\pi}\frac{f_{B_s}^2 m_{B_s}}{\Lambda^2}\bigg[|g_{PS}|^2\frac{m_{B_s}^2}{(m_b + m_s)^2} + \frac{1}{4}|g_{AV}|^2 \frac{m_{B_s}^2}{\Lambda^2
}\bigg]\,,
\end{align}
where $\beta_X = \sqrt{1 - 4m_X^2/m_{B_s}^2}$. Again, $g_{AV}$ vanishes if $\phi = \bar{\phi}$.

The decay rate for outgoing invisible fermions is given by,
\begin{align}
\Gamma(B_s \to \psi\bar{\psi}) = \frac{\beta_\psi}{8\pi}\frac{f_{B_s}^2 m_{B_s}^3}{\Lambda^4}\bigg[&4|f_{AA}|^2\frac{m_\psi^2(m_b+m_s)^2}{m_{B_s}^4} + |f_{PS}|^2 \beta_\psi^2 + |f_{PP}|^2 \nonumber \\
&- 4\,\Re\big[f_{AA}f_{PP}^*\big]\frac{m_\psi(m_b + m_s)}{m_{B_s}^2}\bigg]\frac{m_{B_s}^2}{(m_b+m_s)^2}\,.
\end{align}
Likewise, for a pair of outgoing vector bosons the decay rate is
\begin{align}
\Gamma(B_s \to V\bar{V}) = \frac{\beta_V |h_P|^2}{64\pi}\frac{f_{B_s}^2 m_{B_s}^5}{m_V^4 \Lambda^2}\mathcal{J}_V \frac{m_{B_s}^2}{(m_b + m_s)^2}\,,
\end{align}
with $\mathcal{J}_V = (3 - 2\beta_V^2 +3\beta_V^4)/4$.

Finally, the decay rate for a pair of final state spin 3/2 fermions is
\begin{align}
\Gamma(B_s \to \Psi\bar{\Psi}) = \frac{\beta_\Psi}{72\pi}\frac{f_{B_s}^2 m_{B_s}^7}{m_\Psi^4\Lambda^4}\bigg[& 4|F_{AA}|^2\mathcal{J}_\Psi \frac{m_\Psi^2(m_b + m_s)^2}{m_{B_s}^4} + |F_{PS}|^2\beta_\Psi^2\mathcal{J}_{\Psi}' + |F_{PP}|^2 \mathcal{J}_\Psi \nonumber \\
& - 4\,\Re\big[F_{AA}F_{PP}^*\big]\mathcal{J}_{\Psi}\frac{m_\Psi(m_b+m_s)}{m_{B_s}^2}\bigg]\frac{m_{B_s}^2}{(m_b+m_s)^2}\,,
\end{align}
where the $\mathcal{J}_\Psi$ and $\mathcal{J}_\Psi'$ factors are the same as those below Eq.~\eqref{eq:BK_PsiPsi}. Again, all of the rates above must be multiplied by a factor $1/2$ in the case of identical outgoing states ($\phi = \phi^\dagger$, $\psi = \psi^c$, $V = V^\dagger$, $\Psi = \Psi^c$), and the couplings re-scaled in the same way as the previous sub-section.

\end{widetext}

%%%%%%%%%%%%%%%%%%%%%%%%%%%%%%%%%%%%%%%%%%
\subsection{Standard Model Rates}
%%%%%%%%%%%%%%%%%%%%%%%%%%%%%%%%%%%%%%%%%%

In the SM, the signal $B \to K^{(*)} E_{\text{miss}}$ is induced by the process $B \to K^{(*)}\nu\bar{\nu}$, with the effective Hamiltonian,
\begin{align}
\mathcal{H}_{\text{eff}}^{\text{SM}} = - \frac{4 G_F}{\sqrt{2}}\lambda_t C_L^{\text{SM}}\mathcal{O}_L + \text{h.c.}\,,
\end{align}
where $\lambda_t = V_{tb}V_{ts}^*$ and $C_L^{\text{SM}} = - X_t/s_w^2$, with the factor $X_t =  1.469 \pm 0.017$ accounting for NLO QCD~\cite{Buchalla:1993wq,Misiak:1999yg,Buchalla:1998ba} and two-loop electroweak~\cite{Brod:2010hi} corrections, and the left-handed operator is
\begin{align}
\mathcal{O}_L = \frac{e^2}{16\pi^2}(\bar{b}\gamma_\mu P_L s)(\bar{\nu}_\alpha\gamma^\mu(1-\gamma_5)\nu_\alpha) \,.
\end{align}
In the operator basis of this work, this corresponds to 
\begin{align}
C^{V,LL}_{\underset{sb\alpha\beta}{d\nu}} = \frac{4G_F \lambda_t}{\sqrt{2}}\frac{\alpha}{2\pi}\frac{X_t}{s_w^2}\delta_{\alpha\beta}\,,
\end{align}
which can be inserted into Eqs.~\eqref{eq:chiraltoparity},~\eqref{eq:BK_psipsi} and~\eqref{eq:BKs_psipsi} to obtain the differential rates
\begin{gather}
\label{eq:SM_rates}
\frac{d\Gamma_{\text{SM}}}{dq^2} = \frac{G_F^2 \alpha^2 \left|\lambda_t\right|^2 X_t^2}{32 \pi^5 s_w^4} \left|\vec{p}_{K^{(*)}}\right| q^2 f_{K^{(*)}}^2(q^2)\,,
\end{gather}
where the $K^{(*)}$ three-momentum is given by Eq.~\eqref{eq:K-3momentum} with $m_\phi^2 \to q^2$. 
In Eq.~\eqref{eq:SM_rates}, we have defined the relevant hadronic form factor combinations
\begin{align}
f_K^2(q^2) &\equiv \frac{\left|\vec{p}_{K}\right|^2}{q^2}f_+^2(q^2)\,, \\
f_{K^*}^2(q^2) & \equiv \frac{2 |\vec{p}_{K^*}|^2}{(m_B + m_{K^*})^2}V^{2}(q^2) + \frac{(m_B + m_{K^*})^2}{2m_B^2} A_1^{2}(q^2) \nonumber\\
&\quad  + \frac{16m_{K^*}^2}{q^2} A_{12}^{2}(q^2) \,,  
\end{align}
where the form factors $f_+(q^2)$, $V(q^2)$, $A_1(q^2)$ and $A_{12}(q^2)$ are defined in Appendix~\ref{app:ffs}.

On the other hand, the invisible $B_s$ decay width is negligible in the SM. The two-body $B_s \to \nu \bar \nu$ rate is suppressed by tiny neutrino masses, so the total invisible width is dominated by the four-body decay $B_s \to \nu \bar \nu \nu \bar \nu$, yielding $\mathcal B(B_s \to E_{\rm miss}) \simeq 5 \times 10^{-15} $~\cite{Bhattacharya:2018msv}.

%%%%%%%%%%%%%%%%%%%%%%%%%%%%%%%%%%%%%%%%%%
%
\section{Hadronic Matrix Elements}
\label{app:ffs}
%
%%%%%%%%%%%%%%%%%%%%%%%%%%%%%%%%%%%%%%%%%%

The latest form factors for both the $B\to K$ and $B\to K^*$ transitions can be found in Ref.~\cite{Gubernari:2023puw}, where the authors perform a dispersive analysis of both sets of form factors. The most up-to-date results for the $B\to K$ form factors coming from the lattice are in Ref.~\cite{Parrott:2022rgu}, while in the case of $B\to K^*$ a determination on the lattice of the full set of form factors can be found in Refs.~\cite{Horgan:2013hoa,Horgan:2015vla}, which can be complemented by the LCSR determination in Ref.~\cite{Gubernari:2018wyi}.

\begin{widetext}
%%%%%%%%%%%%%%%%%%%%%%%%%%%%%%%%%%%%%%%%%%
\subsection{$B\to P$ Form Factors} 
%%%%%%%%%%%%%%%%%%%%%%%%%%%%%%%%%%%%%%%%%%

The relevant hadronic matrix elements for the decay of a $B$ meson to a pseudoscalar meson $P$, i.e. $B(p)\to P(k)$, can be written as follows:
\begin{align}
\bra{P}\bar q b\ket{B} &=
\frac{m_B^2-m_P^2}{m_b - m_q}f_0(q^2)\,,
\label{eq:BtoPscalarFF} \\
\bra{P}\bar q \gamma_\mu b \ket{B} &= 
\bigg[P_\mu - \frac{m_B^2-m_P^2}{q^2}q_\mu\bigg]f_+(q^2) + \frac{m_B^2-m_P^2}{q^2} q_\mu f_0(q^2)\,,
\label{eq:BtoPvectorFF} \\
\bra{P}\bar q  \sigma_{\mu\nu} b\ket{B} &=
\frac{i(P_\mu q_\nu - P_\nu q_\mu)}{m_B+m_P}f_T(q^2)\,,
\label{eq:BtoPtensorFF}
\end{align}
where $P_\mu = p_\mu + k_\mu$, $q_\mu = p_\mu - k_\mu$, and the relation $f_+(0)=f_0(0)$ holds. The form factors for the quark currents $\bar{q}\gamma_5 b$ and $\bar{q}\gamma_\mu\gamma_5 b$ vanish identically, while the form factor for $\bar{q}\sigma_{\mu\nu}\gamma_5 b$ can be obtained from the tensor form factor in Eq.~\eqref{eq:BtoPtensorFF} via the identity $\sigma_{\mu\nu}\gamma_5 = \frac{i}{2}\varepsilon_{\mu\nu\alpha\beta}\sigma^{\alpha\beta}$.

%%%%%%%%%%%%%%%%%%%%%%%%%%%%%%%%%%%%%%%%%%
\subsection{$B\to V$ Form Factors} 
%%%%%%%%%%%%%%%%%%%%%%%%%%%%%%%%%%%%%%%%%%

The relevant hadronic matrix elements for $B$ meson decay to a vector meson $V$, i.e. $B(p)\to V(k, \epsilon^*)$, can be expressed as
\begin{align}
\bra{V}\bar{q} \gamma_5 b\ket{B} &= - \frac{2 i m_V \epsilon^{*}_\mu q^\mu}{m_b + m_q} A_0(q^2) \,, \label{eq:BtoVpseudoscalarFF}\\
\bra{V}\bar{q} \gamma_\mu b\ket{B} &=
\frac{2\varepsilon_{\mu\nu\alpha\beta} \epsilon^{*\nu} p^\alpha k^\beta}{m_B + m_V} V(q^2)\,,
\label{eq:BtoVvectorFF} \\
\bra{V}\bar{q} \gamma_\mu \gamma_5 b\ket{B} &=
i \epsilon^{*\nu} \bigg[\frac{2m_V q_\mu q_\nu}{q^2}\big(A_0(q^2) - A_3(q^2)\big) + (m_B + m_V)g_{\mu\nu} A_{1}(q^2) - \frac{P_\mu q_\nu}{m_B + m_V} A_2(q^2)\bigg]\label{eq:BtoVaxialvectorFF} \\
\bra{V}\bar{q} \sigma_{\mu\nu} b\ket{B} &= -i \varepsilon_{\mu\nu\alpha\beta}\epsilon^{*}_{\rho} \bigg[\bigg(P^\alpha  - \frac{m_B^2 - m_V^2}{q^2}q^\alpha\bigg)g^{\beta\rho} T_{1}(q^2) + \frac{m_B^2 - m_V^2}{q^2}q^\alpha g^{\beta\rho} T_{2}(q^2) \nonumber \\
&\hspace{6em} + \frac{q^\alpha  P^\beta}{q^2} q^\rho\big(T_{1}(q^2) - \widetilde{T}_{3}(q^2)\big)\bigg]\,,\label{eq:BtoVtensorFF}
\end{align}
\end{widetext}
where $A_0(0) = A_3(0)$, $T_1(0) = T_2(0)$, and
\begin{align}
\widetilde{T}_{3}(q^2) \equiv T_{2}(q^2) + \frac{q^2}{m_B^2 - m_V^2}T_{3}(q^2)\,.
\end{align}
The form factor for the scalar quark current $\bar{b}q$ vanishes identically, while the form factor for $\bar{b}\sigma_{\mu\nu}\gamma_5 q$ can again be obtained from Eq.~\eqref{eq:BtoVtensorFF} via $\sigma_{\mu\nu}\gamma_5 = \frac{i}{2}\varepsilon_{\mu\nu\alpha\beta}\sigma^{\alpha\beta}$.

The form factor $A_{3}(q^2)$ can be eliminated using the relation
\begin{align}
A_3(q^2) = \frac{m_B+m_V}{2m_V} A_1(q^2) -\frac{m_B-m_V}{2m_V} A_2(q^2)\,.
\end{align}
The form factors $A_2(q^2)$ and $T_{3}(q^2)$ can likewise be eliminated in favour of the so-called helicity form factors,
\begin{widetext}
\begin{align}
A_{12}(q^2) &\equiv \frac{(m_B + m_V)^2(m_B^2 - m_V^2 - q^2)A_{1}(q^2) - \lambda(m_B^2, q^2, m_V^2) A_{2}(q^2)}{16 m_B m_V^2 (m_B + m_V)} \,, \\
T_{23}(q^2) &\equiv \frac{(m_B^2 - m_V^2)(m_B^2 + 3 m_V^2 - q^2)T_{2}(q^2) - \lambda(m_B^2, q^2, m_V^2) T_{3}(q^2)}{8 m_B m_V^2 (m_B - m_V)}\,,
\end{align}
which appear in decay rates for the longitudinal polarisation of the vector meson. 
\end{widetext}

The required form factors for $B\to K^{(*)}$ decays are therefore $f_+$, $f_0$, $f_T$, $V$, $A_0$, $A_1$, $A_{12}$, $T_{1}$, $T_{2}$ and $T_{23}$. For these, we use the BSZ parametrisation~\cite{Bharucha:2015bzk} results of Ref.~\cite{Gubernari:2023puw}.

%%%%%%%%%%%%%%%%%%%%%%%%%%%%%%%%%%%%%%%%%%
\subsection{$B_s$ Decay Matrix Elements}
\label{Bsffs}
%%%%%%%%%%%%%%%%%%%%%%%%%%%%%%%%%%%%%%%%%%

The relevant matrix elements for $B_s$ decay, which are used to derive the decay rates $B_s\to\phi\bar{\phi}/\psi\bar{\psi}/V\bar{V}/\Psi\bar{\Psi}$ in Appendix~\ref{app:decays}, are
\begin{align}
\bra{0}\bar{s} \gamma_\mu \gamma_5 b\ket{B_s} = i f_{B_s}P_\mu \,, \nonumber \\
\bra{0}\bar{s} \gamma_5 b\ket{B_s} = - i \frac{m_{B_s}^2 f_{B_s}}{m_b + m_s} \,,
\end{align}
where $P_\mu$ is the four-momentum of $B_s$ and we take the value $f_{B_s} = 230.3(1.3)~\text{MeV}$ for the $B_s$ decay constant~\cite{FlavourLatticeAveragingGroupFLAG:2021npn}.

%%%%%%%%%%%%%%%%%%%%%%%%%%%%%%%%%%%%%%%%%%
%
\section{Experimental Likelihood Reconstruction}
\label{app:ExpLikelihoodReconstruction}
%
%%%%%%%%%%%%%%%%%%%%%%%%%%%%%%%%%%%%%%%%%%

In this appendix, we give further details on the reconstruction of the likelihoods for the Belle~II and BaBar analyses~\cite{Belle-II:2023esi,BaBar:2013npw}.

The Belle~II~\cite{Belle-II:2023esi} analysis measured the differential decay width for the charged channel $B^{+}\to K^{+} E_{\rm miss}$ with both inclusive (ITA) and hadronic (HTA) tag methods. In order to calculate the expected signal events using Eq.~\eqref{eq:exp_sig}, we have been provided with the smearing $f_{q^2_{\rm rec}}(q^2)$ of the reconstructed momentum transfer $q^2_{\rm rec}$ for the ITA analysis, and with the efficiencies for both ITA and HTA analyses \cite{Belle-II:PrivateCommunication}. 
We computed the Monte-Carlo statistical errors for each background component ($B^{+}B^{-}$, $B^{0}\bar{B}^{0}$ and continuum) as $\sigma^{i}_{b, {\rm stat}}=\sqrt{b^i/F_{\rm MC}}$, where $b^i$ corresponds to the number of background events in the bin $i$ and $F_{\rm MC}$ is a bin independent Monte-Carlo factor that reproduces the simulation statistical uncertainties in Fig.~17 of Ref.~\cite{Belle-II:2023esi}. As discussed in the main text, we performed a Monte-Carlo simulation of the expected SM signal including uncertainties on the efficiencies and form factors, to find the covariance $\Sigma_{\text{SM}}$, including correlations between bins. The covariances for the background components were then found by re-scaling $\Sigma_{\text{SM}}$ according to the background statistical errors $\sigma^{i}_{b, {\rm stat}}$.
We furthermore introduce a systematic uncertainty for the normalisation of the backgrounds, which are fitted to reproduce the profile log-likelihood ratio as a function of the signal strength given in Fig. 16 of Ref.~\cite{Belle-II:2023esi}.

The BaBar~\cite{BaBar:2013npw} analysis performed a search for both the neutral and charged channels $B^{0(+)}\to K^{*0(+)} E_{\rm miss}$, combining two different final states for each channel ($K^+\pi^-$, $K_S\pi^0$, $K^+\pi^0$ and $K_S\pi^+$) and using a hadronic tag method. We extract the different background contributions and their $q^2$-dependence from Fig.~5 of Ref.~\cite{BaBar:2013npw}, while we extract the efficiencies $\epsilon(q^2)$ from Fig.~6 in their binned forms. We compute the statistical errors similarly to the Belle~II analysis, fitting the Monte-Carlo factor to reproduce the values in Table~IV of Ref.~\cite{BaBar:2013npw}. We obtained the systematic uncertainties from Table~II and III  of Ref.~\cite{BaBar:2013npw}, taking into account the correlation between the systematic uncertainties in the background and efficiency.

Since no information is available on the smearing $f_{q^2_{\rm rec}}(q^2)$ of the reconstructed momentum transfer $q^2_{\rm rec}$, we exclude the constraints obtained from $B^{+}\to K^{*+} E_{\rm miss}$, as both final states contain neutral particles leading to non-negligible and asymmetrical smearing. In the case $B^{0}\to K^{*0} E_{\rm miss}$ we assume that the $K^{*0}$ dominantly decays to  $K^+\pi^-$ (as there is a factor of 10 in the efficiency) and we therefore neglect the smearing.   

Lastly, we have verified that our likelihoods reproduce correctly the confidence intervals and limits given in both Refs.~\cite{Belle-II:2023esi,BaBar:2013npw}.

%%%%%%%%%%%%%%%%%%%%%%%%%%%%%%%%%%%%%%%%%%

%\nocite{*}
 
\bibliography{main}

\begin{thebibliography}{54}%
\makeatletter
\providecommand \@ifxundefined [1]{%
 \@ifx{#1\undefined}
}%
\providecommand \@ifnum [1]{%
 \ifnum #1\expandafter \@firstoftwo
 \else \expandafter \@secondoftwo
 \fi
}%
\providecommand \@ifx [1]{%
 \ifx #1\expandafter \@firstoftwo
 \else \expandafter \@secondoftwo
 \fi
}%
\providecommand \natexlab [1]{#1}%
\providecommand \enquote  [1]{``#1''}%
\providecommand \bibnamefont  [1]{#1}%
\providecommand \bibfnamefont [1]{#1}%
\providecommand \citenamefont [1]{#1}%
\providecommand \href@noop [0]{\@secondoftwo}%
\providecommand \href [0]{\begingroup \@sanitize@url \@href}%
\providecommand \@href[1]{\@@startlink{#1}\@@href}%
\providecommand \@@href[1]{\endgroup#1\@@endlink}%
\providecommand \@sanitize@url [0]{\catcode `\\12\catcode `\$12\catcode
  `\&12\catcode `\#12\catcode `\^12\catcode `\_12\catcode `\%12\relax}%
\providecommand \@@startlink[1]{}%
\providecommand \@@endlink[0]{}%
\providecommand \url  [0]{\begingroup\@sanitize@url \@url }%
\providecommand \@url [1]{\endgroup\@href {#1}{\urlprefix }}%
\providecommand \urlprefix  [0]{URL }%
\providecommand \Eprint [0]{\href }%
\providecommand \doibase [0]{http://dx.doi.org/}%
\providecommand \selectlanguage [0]{\@gobble}%
\providecommand \bibinfo  [0]{\@secondoftwo}%
\providecommand \bibfield  [0]{\@secondoftwo}%
\providecommand \translation [1]{[#1]}%
\providecommand \BibitemOpen [0]{}%
\providecommand \bibitemStop [0]{}%
\providecommand \bibitemNoStop [0]{.\EOS\space}%
\providecommand \EOS [0]{\spacefactor3000\relax}%
\providecommand \BibitemShut  [1]{\csname bibitem#1\endcsname}%
\let\auto@bib@innerbib\@empty
%</preamble>
\bibitem [{\citenamefont {Amhis}\ \emph {et~al.}(2023)\citenamefont {Amhis}
  \emph {et~al.}}]{HFLAV:2022esi}%
  \BibitemOpen
  \bibfield  {author} {\bibinfo {author} {\bibfnamefont {Y.~S.}\ \bibnamefont
  {Amhis}} \emph {et~al.} (\bibinfo {collaboration} {HFLAV}),\ }\href {\doibase
  10.1103/PhysRevD.107.052008} {\bibfield  {journal} {\bibinfo  {journal}
  {Phys. Rev. D}\ }\textbf {\bibinfo {volume} {107}},\ \bibinfo {pages}
  {052008} (\bibinfo {year} {2023})},\ \Eprint
  {http://arxiv.org/abs/2206.07501} {arXiv:2206.07501 [hep-ex]} \BibitemShut
  {NoStop}%
\bibitem [{\citenamefont {Alguer\'o}\ \emph {et~al.}(2023)\citenamefont
  {Alguer\'o}, \citenamefont {Biswas}, \citenamefont {Capdevila}, \citenamefont
  {Descotes-Genon}, \citenamefont {Matias},\ and\ \citenamefont
  {Novoa-Brunet}}]{Alguero:2023jeh}%
  \BibitemOpen
  \bibfield  {author} {\bibinfo {author} {\bibfnamefont {M.}~\bibnamefont
  {Alguer\'o}}, \bibinfo {author} {\bibfnamefont {A.}~\bibnamefont {Biswas}},
  \bibinfo {author} {\bibfnamefont {B.}~\bibnamefont {Capdevila}}, \bibinfo
  {author} {\bibfnamefont {S.}~\bibnamefont {Descotes-Genon}}, \bibinfo
  {author} {\bibfnamefont {J.}~\bibnamefont {Matias}}, \ and\ \bibinfo {author}
  {\bibfnamefont {M.}~\bibnamefont {Novoa-Brunet}},\ }\href {\doibase
  10.1140/epjc/s10052-023-11824-0} {\bibfield  {journal} {\bibinfo  {journal}
  {Eur. Phys. J. C}\ }\textbf {\bibinfo {volume} {83}},\ \bibinfo {pages} {648}
  (\bibinfo {year} {2023})},\ \Eprint {http://arxiv.org/abs/2304.07330}
  {arXiv:2304.07330 [hep-ph]} \BibitemShut {NoStop}%
\bibitem [{\citenamefont {Adachi}\ \emph {et~al.}(2023)\citenamefont {Adachi}
  \emph {et~al.}}]{Belle-II:2023esi}%
  \BibitemOpen
  \bibfield  {author} {\bibinfo {author} {\bibfnamefont {I.}~\bibnamefont
  {Adachi}} \emph {et~al.} (\bibinfo {collaboration} {Belle-II}),\ }\href@noop
  {} {\  (\bibinfo {year} {2023})},\ \Eprint {http://arxiv.org/abs/2311.14647}
  {arXiv:2311.14647 [hep-ex]} \BibitemShut {NoStop}%
\bibitem [{\citenamefont {Lees}\ \emph {et~al.}(2013)\citenamefont {Lees} \emph
  {et~al.}}]{BaBar:2013npw}%
  \BibitemOpen
  \bibfield  {author} {\bibinfo {author} {\bibfnamefont {J.~P.}\ \bibnamefont
  {Lees}} \emph {et~al.} (\bibinfo {collaboration} {BaBar}),\ }\href {\doibase
  10.1103/PhysRevD.87.112005} {\bibfield  {journal} {\bibinfo  {journal} {Phys.
  Rev. D}\ }\textbf {\bibinfo {volume} {87}},\ \bibinfo {pages} {112005}
  (\bibinfo {year} {2013})},\ \Eprint {http://arxiv.org/abs/1303.7465}
  {arXiv:1303.7465 [hep-ex]} \BibitemShut {NoStop}%
\bibitem [{\citenamefont {Alonso-\'Alvarez}\ and\ \citenamefont
  {Escudero}(2023)}]{Alonso-Alvarez:2023mgc}%
  \BibitemOpen
  \bibfield  {author} {\bibinfo {author} {\bibfnamefont {G.}~\bibnamefont
  {Alonso-\'Alvarez}}\ and\ \bibinfo {author} {\bibfnamefont {M.}~\bibnamefont
  {Escudero}},\ }\href@noop {} {\  (\bibinfo {year} {2023})},\ \Eprint
  {http://arxiv.org/abs/2310.13043} {arXiv:2310.13043 [hep-ph]} \BibitemShut
  {NoStop}%
\bibitem [{\citenamefont {Barate}\ \emph {et~al.}(2001)\citenamefont {Barate}
  \emph {et~al.}}]{ALEPH:2000vvi}%
  \BibitemOpen
  \bibfield  {author} {\bibinfo {author} {\bibfnamefont {R.}~\bibnamefont
  {Barate}} \emph {et~al.} (\bibinfo {collaboration} {ALEPH}),\ }\href
  {\doibase 10.1007/s100520100612} {\bibfield  {journal} {\bibinfo  {journal}
  {Eur. Phys. J. C}\ }\textbf {\bibinfo {volume} {19}},\ \bibinfo {pages} {213}
  (\bibinfo {year} {2001})},\ \Eprint {http://arxiv.org/abs/hep-ex/0010022}
  {arXiv:hep-ex/0010022} \BibitemShut {NoStop}%
\bibitem [{\citenamefont {Buras}(2020)}]{Buras:2020xsm}%
  \BibitemOpen
  \bibfield  {author} {\bibinfo {author} {\bibfnamefont {A.}~\bibnamefont
  {Buras}},\ }\href {\doibase 10.1017/9781139524100} {\emph {\bibinfo {title}
  {{Gauge Theory of Weak Decays}}}}\ (\bibinfo  {publisher} {Cambridge
  University Press},\ \bibinfo {year} {2020})\BibitemShut {NoStop}%
\bibitem [{\citenamefont {Buras}\ \emph {et~al.}(2015)\citenamefont {Buras},
  \citenamefont {Girrbach-Noe}, \citenamefont {Niehoff},\ and\ \citenamefont
  {Straub}}]{Buras:2014fpa}%
  \BibitemOpen
  \bibfield  {author} {\bibinfo {author} {\bibfnamefont {A.~J.}\ \bibnamefont
  {Buras}}, \bibinfo {author} {\bibfnamefont {J.}~\bibnamefont {Girrbach-Noe}},
  \bibinfo {author} {\bibfnamefont {C.}~\bibnamefont {Niehoff}}, \ and\
  \bibinfo {author} {\bibfnamefont {D.~M.}\ \bibnamefont {Straub}},\ }\href
  {\doibase 10.1007/JHEP02(2015)184} {\bibfield  {journal} {\bibinfo  {journal}
  {JHEP}\ }\textbf {\bibinfo {volume} {02}},\ \bibinfo {pages} {184} (\bibinfo
  {year} {2015})},\ \Eprint {http://arxiv.org/abs/1409.4557} {arXiv:1409.4557
  [hep-ph]} \BibitemShut {NoStop}%
\bibitem [{\citenamefont {Be\v{c}irevi\'c}\ \emph {et~al.}(2023)\citenamefont
  {Be\v{c}irevi\'c}, \citenamefont {Piazza},\ and\ \citenamefont
  {Sumensari}}]{Becirevic:2023aov}%
  \BibitemOpen
  \bibfield  {author} {\bibinfo {author} {\bibfnamefont {D.}~\bibnamefont
  {Be\v{c}irevi\'c}}, \bibinfo {author} {\bibfnamefont {G.}~\bibnamefont
  {Piazza}}, \ and\ \bibinfo {author} {\bibfnamefont {O.}~\bibnamefont
  {Sumensari}},\ }\href {\doibase 10.1140/epjc/s10052-023-11388-z} {\bibfield
  {journal} {\bibinfo  {journal} {Eur. Phys. J. C}\ }\textbf {\bibinfo {volume}
  {83}},\ \bibinfo {pages} {252} (\bibinfo {year} {2023})},\ \Eprint
  {http://arxiv.org/abs/2301.06990} {arXiv:2301.06990 [hep-ph]} \BibitemShut
  {NoStop}%
\bibitem [{\citenamefont {Gubernari}\ \emph {et~al.}(2023)\citenamefont
  {Gubernari}, \citenamefont {Reboud}, \citenamefont {van Dyk},\ and\
  \citenamefont {Virto}}]{Gubernari:2023puw}%
  \BibitemOpen
  \bibfield  {author} {\bibinfo {author} {\bibfnamefont {N.}~\bibnamefont
  {Gubernari}}, \bibinfo {author} {\bibfnamefont {M.}~\bibnamefont {Reboud}},
  \bibinfo {author} {\bibfnamefont {D.}~\bibnamefont {van Dyk}}, \ and\
  \bibinfo {author} {\bibfnamefont {J.}~\bibnamefont {Virto}},\ }\href
  {\doibase 10.1007/JHEP12(2023)153} {\bibfield  {journal} {\bibinfo  {journal}
  {JHEP}\ }\textbf {\bibinfo {volume} {12}},\ \bibinfo {pages} {153} (\bibinfo
  {year} {2023})},\ \Eprint {http://arxiv.org/abs/2305.06301} {arXiv:2305.06301
  [hep-ph]} \BibitemShut {NoStop}%
\bibitem [{\citenamefont {Athron}\ \emph {et~al.}(2023)\citenamefont {Athron},
  \citenamefont {Martinez},\ and\ \citenamefont {Sierra}}]{Athron:2023hmz}%
  \BibitemOpen
  \bibfield  {author} {\bibinfo {author} {\bibfnamefont {P.}~\bibnamefont
  {Athron}}, \bibinfo {author} {\bibfnamefont {R.}~\bibnamefont {Martinez}}, \
  and\ \bibinfo {author} {\bibfnamefont {C.}~\bibnamefont {Sierra}},\
  }\href@noop {} {\  (\bibinfo {year} {2023})},\ \Eprint
  {http://arxiv.org/abs/2308.13426} {arXiv:2308.13426 [hep-ph]} \BibitemShut
  {NoStop}%
\bibitem [{\citenamefont {Bause}\ \emph {et~al.}(2024)\citenamefont {Bause},
  \citenamefont {Gisbert},\ and\ \citenamefont {Hiller}}]{Bause:2023mfe}%
  \BibitemOpen
  \bibfield  {author} {\bibinfo {author} {\bibfnamefont {R.}~\bibnamefont
  {Bause}}, \bibinfo {author} {\bibfnamefont {H.}~\bibnamefont {Gisbert}}, \
  and\ \bibinfo {author} {\bibfnamefont {G.}~\bibnamefont {Hiller}},\ }\href
  {\doibase 10.1103/PhysRevD.109.015006} {\bibfield  {journal} {\bibinfo
  {journal} {Phys. Rev. D}\ }\textbf {\bibinfo {volume} {109}},\ \bibinfo
  {pages} {015006} (\bibinfo {year} {2024})},\ \Eprint
  {http://arxiv.org/abs/2309.00075} {arXiv:2309.00075 [hep-ph]} \BibitemShut
  {NoStop}%
\bibitem [{\citenamefont {Allwicher}\ \emph {et~al.}(2024)\citenamefont
  {Allwicher}, \citenamefont {Becirevic}, \citenamefont {Piazza}, \citenamefont
  {Rosauro-Alcaraz},\ and\ \citenamefont {Sumensari}}]{Allwicher:2023xba}%
  \BibitemOpen
  \bibfield  {author} {\bibinfo {author} {\bibfnamefont {L.}~\bibnamefont
  {Allwicher}}, \bibinfo {author} {\bibfnamefont {D.}~\bibnamefont
  {Becirevic}}, \bibinfo {author} {\bibfnamefont {G.}~\bibnamefont {Piazza}},
  \bibinfo {author} {\bibfnamefont {S.}~\bibnamefont {Rosauro-Alcaraz}}, \ and\
  \bibinfo {author} {\bibfnamefont {O.}~\bibnamefont {Sumensari}},\ }\href
  {\doibase 10.1016/j.physletb.2023.138411} {\bibfield  {journal} {\bibinfo
  {journal} {Phys. Lett. B}\ }\textbf {\bibinfo {volume} {848}},\ \bibinfo
  {pages} {138411} (\bibinfo {year} {2024})},\ \Eprint
  {http://arxiv.org/abs/2309.02246} {arXiv:2309.02246 [hep-ph]} \BibitemShut
  {NoStop}%
\bibitem [{\citenamefont {Chen}\ and\ \citenamefont
  {Chiang}(2023)}]{Chen:2023wpb}%
  \BibitemOpen
  \bibfield  {author} {\bibinfo {author} {\bibfnamefont {C.-H.}\ \bibnamefont
  {Chen}}\ and\ \bibinfo {author} {\bibfnamefont {C.-W.}\ \bibnamefont
  {Chiang}},\ }\href@noop {} {\  (\bibinfo {year} {2023})},\ \Eprint
  {http://arxiv.org/abs/2309.12904} {arXiv:2309.12904 [hep-ph]} \BibitemShut
  {NoStop}%
\bibitem [{\citenamefont {Felkl}\ \emph {et~al.}(2023)\citenamefont {Felkl},
  \citenamefont {Giri}, \citenamefont {Mohanta},\ and\ \citenamefont
  {Schmidt}}]{Felkl:2023ayn}%
  \BibitemOpen
  \bibfield  {author} {\bibinfo {author} {\bibfnamefont {T.}~\bibnamefont
  {Felkl}}, \bibinfo {author} {\bibfnamefont {A.}~\bibnamefont {Giri}},
  \bibinfo {author} {\bibfnamefont {R.}~\bibnamefont {Mohanta}}, \ and\
  \bibinfo {author} {\bibfnamefont {M.~A.}\ \bibnamefont {Schmidt}},\ }\href
  {\doibase 10.1140/epjc/s10052-023-12326-9} {\bibfield  {journal} {\bibinfo
  {journal} {Eur. Phys. J. C}\ }\textbf {\bibinfo {volume} {83}},\ \bibinfo
  {pages} {1135} (\bibinfo {year} {2023})},\ \Eprint
  {http://arxiv.org/abs/2309.02940} {arXiv:2309.02940 [hep-ph]} \BibitemShut
  {NoStop}%
\bibitem [{\citenamefont {Abdughani}\ and\ \citenamefont
  {Reyimuaji}(2023)}]{Abdughani:2023dlr}%
  \BibitemOpen
  \bibfield  {author} {\bibinfo {author} {\bibfnamefont {M.}~\bibnamefont
  {Abdughani}}\ and\ \bibinfo {author} {\bibfnamefont {Y.}~\bibnamefont
  {Reyimuaji}},\ }\href@noop {} {\  (\bibinfo {year} {2023})},\ \Eprint
  {http://arxiv.org/abs/2309.03706} {arXiv:2309.03706 [hep-ph]} \BibitemShut
  {NoStop}%
\bibitem [{\citenamefont {Dreiner}\ \emph {et~al.}(2023)\citenamefont
  {Dreiner}, \citenamefont {G\"unther},\ and\ \citenamefont
  {Wang}}]{Dreiner:2023cms}%
  \BibitemOpen
  \bibfield  {author} {\bibinfo {author} {\bibfnamefont {H.~K.}\ \bibnamefont
  {Dreiner}}, \bibinfo {author} {\bibfnamefont {J.~Y.}\ \bibnamefont
  {G\"unther}}, \ and\ \bibinfo {author} {\bibfnamefont {Z.~S.}\ \bibnamefont
  {Wang}},\ }\href@noop {} {\  (\bibinfo {year} {2023})},\ \Eprint
  {http://arxiv.org/abs/2309.03727} {arXiv:2309.03727 [hep-ph]} \BibitemShut
  {NoStop}%
\bibitem [{\citenamefont {He}\ \emph {et~al.}(2023)\citenamefont {He},
  \citenamefont {Ma},\ and\ \citenamefont {Valencia}}]{He:2023bnk}%
  \BibitemOpen
  \bibfield  {author} {\bibinfo {author} {\bibfnamefont {X.-G.}\ \bibnamefont
  {He}}, \bibinfo {author} {\bibfnamefont {X.-D.}\ \bibnamefont {Ma}}, \ and\
  \bibinfo {author} {\bibfnamefont {G.}~\bibnamefont {Valencia}},\ }\href@noop
  {} {\  (\bibinfo {year} {2023})},\ \Eprint {http://arxiv.org/abs/2309.12741}
  {arXiv:2309.12741 [hep-ph]} \BibitemShut {NoStop}%
\bibitem [{\citenamefont {Berezhnoy}\ and\ \citenamefont
  {Melikhov}(2023)}]{Berezhnoy:2023rxx}%
  \BibitemOpen
  \bibfield  {author} {\bibinfo {author} {\bibfnamefont {A.}~\bibnamefont
  {Berezhnoy}}\ and\ \bibinfo {author} {\bibfnamefont {D.}~\bibnamefont
  {Melikhov}},\ }\href@noop {} {\  (\bibinfo {year} {2023})},\ \Eprint
  {http://arxiv.org/abs/2309.17191} {arXiv:2309.17191 [hep-ph]} \BibitemShut
  {NoStop}%
\bibitem [{\citenamefont {Datta}\ \emph {et~al.}(2023)\citenamefont {Datta},
  \citenamefont {Marfatia},\ and\ \citenamefont {Mukherjee}}]{Datta:2023iln}%
  \BibitemOpen
  \bibfield  {author} {\bibinfo {author} {\bibfnamefont {A.}~\bibnamefont
  {Datta}}, \bibinfo {author} {\bibfnamefont {D.}~\bibnamefont {Marfatia}}, \
  and\ \bibinfo {author} {\bibfnamefont {L.}~\bibnamefont {Mukherjee}},\
  }\href@noop {} {\  (\bibinfo {year} {2023})},\ \Eprint
  {http://arxiv.org/abs/2310.15136} {arXiv:2310.15136 [hep-ph]} \BibitemShut
  {NoStop}%
\bibitem [{\citenamefont {Altmannshofer}\ \emph {et~al.}(2023)\citenamefont
  {Altmannshofer}, \citenamefont {Crivellin}, \citenamefont {Haigh},
  \citenamefont {Inguglia},\ and\ \citenamefont
  {Martin~Camalich}}]{Altmannshofer:2023hkn}%
  \BibitemOpen
  \bibfield  {author} {\bibinfo {author} {\bibfnamefont {W.}~\bibnamefont
  {Altmannshofer}}, \bibinfo {author} {\bibfnamefont {A.}~\bibnamefont
  {Crivellin}}, \bibinfo {author} {\bibfnamefont {H.}~\bibnamefont {Haigh}},
  \bibinfo {author} {\bibfnamefont {G.}~\bibnamefont {Inguglia}}, \ and\
  \bibinfo {author} {\bibfnamefont {J.}~\bibnamefont {Martin~Camalich}},\
  }\href@noop {} {\  (\bibinfo {year} {2023})},\ \Eprint
  {http://arxiv.org/abs/2311.14629} {arXiv:2311.14629 [hep-ph]} \BibitemShut
  {NoStop}%
\bibitem [{\citenamefont {Fridell}\ \emph {et~al.}(2023)\citenamefont
  {Fridell}, \citenamefont {Ghosh}, \citenamefont {Okui},\ and\ \citenamefont
  {Tobioka}}]{Fridell:2023ssf}%
  \BibitemOpen
  \bibfield  {author} {\bibinfo {author} {\bibfnamefont {K.}~\bibnamefont
  {Fridell}}, \bibinfo {author} {\bibfnamefont {M.}~\bibnamefont {Ghosh}},
  \bibinfo {author} {\bibfnamefont {T.}~\bibnamefont {Okui}}, \ and\ \bibinfo
  {author} {\bibfnamefont {K.}~\bibnamefont {Tobioka}},\ }\href@noop {} {\
  (\bibinfo {year} {2023})},\ \Eprint {http://arxiv.org/abs/2312.12507}
  {arXiv:2312.12507 [hep-ph]} \BibitemShut {NoStop}%
\bibitem [{\citenamefont {Gabrielli}\ \emph {et~al.}(2024)\citenamefont
  {Gabrielli}, \citenamefont {Marzola}, \citenamefont {M\"u\"ursepp},\ and\
  \citenamefont {Raidal}}]{Gabrielli:2024wys}%
  \BibitemOpen
  \bibfield  {author} {\bibinfo {author} {\bibfnamefont {E.}~\bibnamefont
  {Gabrielli}}, \bibinfo {author} {\bibfnamefont {L.}~\bibnamefont {Marzola}},
  \bibinfo {author} {\bibfnamefont {K.}~\bibnamefont {M\"u\"ursepp}}, \ and\
  \bibinfo {author} {\bibfnamefont {M.}~\bibnamefont {Raidal}},\ }\href@noop {}
  {\  (\bibinfo {year} {2024})},\ \Eprint {http://arxiv.org/abs/2402.05901}
  {arXiv:2402.05901 [hep-ph]} \BibitemShut {NoStop}%
\bibitem [{\citenamefont {Chen}\ \emph {et~al.}(2024)\citenamefont {Chen},
  \citenamefont {Wen},\ and\ \citenamefont {Xu}}]{Chen:2024jlj}%
  \BibitemOpen
  \bibfield  {author} {\bibinfo {author} {\bibfnamefont {F.-Z.}\ \bibnamefont
  {Chen}}, \bibinfo {author} {\bibfnamefont {Q.}~\bibnamefont {Wen}}, \ and\
  \bibinfo {author} {\bibfnamefont {F.}~\bibnamefont {Xu}},\ }\href@noop {} {\
  (\bibinfo {year} {2024})},\ \Eprint {http://arxiv.org/abs/2401.11552}
  {arXiv:2401.11552 [hep-ph]} \BibitemShut {NoStop}%
\bibitem [{\citenamefont {Hou}\ \emph {et~al.}(2024)\citenamefont {Hou},
  \citenamefont {Li}, \citenamefont {Shen}, \citenamefont {Yang},\ and\
  \citenamefont {Yuan}}]{Hou:2024vyw}%
  \BibitemOpen
  \bibfield  {author} {\bibinfo {author} {\bibfnamefont {B.-F.}\ \bibnamefont
  {Hou}}, \bibinfo {author} {\bibfnamefont {X.-Q.}\ \bibnamefont {Li}},
  \bibinfo {author} {\bibfnamefont {M.}~\bibnamefont {Shen}}, \bibinfo {author}
  {\bibfnamefont {Y.-D.}\ \bibnamefont {Yang}}, \ and\ \bibinfo {author}
  {\bibfnamefont {X.-B.}\ \bibnamefont {Yuan}},\ }\href@noop {} {\  (\bibinfo
  {year} {2024})},\ \Eprint {http://arxiv.org/abs/2402.19208} {arXiv:2402.19208
  [hep-ph]} \BibitemShut {NoStop}%
\bibitem [{\citenamefont {Ho}\ \emph {et~al.}(2024)\citenamefont {Ho},
  \citenamefont {Kim},\ and\ \citenamefont {Ko}}]{Ho:2024cwk}%
  \BibitemOpen
  \bibfield  {author} {\bibinfo {author} {\bibfnamefont {S.-Y.}\ \bibnamefont
  {Ho}}, \bibinfo {author} {\bibfnamefont {J.}~\bibnamefont {Kim}}, \ and\
  \bibinfo {author} {\bibfnamefont {P.}~\bibnamefont {Ko}},\ }\href@noop {} {\
  (\bibinfo {year} {2024})},\ \Eprint {http://arxiv.org/abs/2401.10112}
  {arXiv:2401.10112 [hep-ph]} \BibitemShut {NoStop}%
\bibitem [{\citenamefont {Descotes-Genon}\ \emph {et~al.}(2020)\citenamefont
  {Descotes-Genon}, \citenamefont {Fajfer}, \citenamefont {Kamenik},\ and\
  \citenamefont {Novoa-Brunet}}]{Descotes-Genon:2020buf}%
  \BibitemOpen
  \bibfield  {author} {\bibinfo {author} {\bibfnamefont {S.}~\bibnamefont
  {Descotes-Genon}}, \bibinfo {author} {\bibfnamefont {S.}~\bibnamefont
  {Fajfer}}, \bibinfo {author} {\bibfnamefont {J.~F.}\ \bibnamefont {Kamenik}},
  \ and\ \bibinfo {author} {\bibfnamefont {M.}~\bibnamefont {Novoa-Brunet}},\
  }\href {\doibase 10.1016/j.physletb.2020.135769} {\bibfield  {journal}
  {\bibinfo  {journal} {Phys. Lett. B}\ }\textbf {\bibinfo {volume} {809}},\
  \bibinfo {pages} {135769} (\bibinfo {year} {2020})},\ \bibinfo {note}
  {[Addendum: Phys.Lett.B 840, 137830 (2023)]},\ \Eprint
  {http://arxiv.org/abs/2005.03734} {arXiv:2005.03734 [hep-ph]} \BibitemShut
  {NoStop}%
\bibitem [{\citenamefont {Kamenik}\ and\ \citenamefont
  {Smith}(2012)}]{Kamenik:2011vy}%
  \BibitemOpen
  \bibfield  {author} {\bibinfo {author} {\bibfnamefont {J.~F.}\ \bibnamefont
  {Kamenik}}\ and\ \bibinfo {author} {\bibfnamefont {C.}~\bibnamefont
  {Smith}},\ }\href {\doibase 10.1007/JHEP03(2012)090} {\bibfield  {journal}
  {\bibinfo  {journal} {JHEP}\ }\textbf {\bibinfo {volume} {03}},\ \bibinfo
  {pages} {090} (\bibinfo {year} {2012})},\ \Eprint
  {http://arxiv.org/abs/1111.6402} {arXiv:1111.6402 [hep-ph]} \BibitemShut
  {NoStop}%
\bibitem [{\citenamefont {Dror}\ \emph {et~al.}(2017)\citenamefont {Dror},
  \citenamefont {Lasenby},\ and\ \citenamefont {Pospelov}}]{Dror:2017nsg}%
  \BibitemOpen
  \bibfield  {author} {\bibinfo {author} {\bibfnamefont {J.~A.}\ \bibnamefont
  {Dror}}, \bibinfo {author} {\bibfnamefont {R.}~\bibnamefont {Lasenby}}, \
  and\ \bibinfo {author} {\bibfnamefont {M.}~\bibnamefont {Pospelov}},\ }\href
  {\doibase 10.1103/PhysRevD.96.075036} {\bibfield  {journal} {\bibinfo
  {journal} {Phys. Rev. D}\ }\textbf {\bibinfo {volume} {96}},\ \bibinfo
  {pages} {075036} (\bibinfo {year} {2017})},\ \Eprint
  {http://arxiv.org/abs/1707.01503} {arXiv:1707.01503 [hep-ph]} \BibitemShut
  {NoStop}%
\bibitem [{\citenamefont {Di~Luzio}\ \emph {et~al.}(2022)\citenamefont
  {Di~Luzio}, \citenamefont {Nardecchia},\ and\ \citenamefont
  {Toni}}]{DiLuzio:2022ziu}%
  \BibitemOpen
  \bibfield  {author} {\bibinfo {author} {\bibfnamefont {L.}~\bibnamefont
  {Di~Luzio}}, \bibinfo {author} {\bibfnamefont {M.}~\bibnamefont
  {Nardecchia}}, \ and\ \bibinfo {author} {\bibfnamefont {C.}~\bibnamefont
  {Toni}},\ }\href {\doibase 10.1103/PhysRevD.105.115042} {\bibfield  {journal}
  {\bibinfo  {journal} {Phys. Rev. D}\ }\textbf {\bibinfo {volume} {105}},\
  \bibinfo {pages} {115042} (\bibinfo {year} {2022})},\ \Eprint
  {http://arxiv.org/abs/2204.05945} {arXiv:2204.05945 [hep-ph]} \BibitemShut
  {NoStop}%
\bibitem [{\citenamefont {Dembinski}\ and\ \citenamefont
  {et~al.}(2020)}]{iminuit}%
  \BibitemOpen
  \bibfield  {author} {\bibinfo {author} {\bibfnamefont {H.}~\bibnamefont
  {Dembinski}}\ and\ \bibinfo {author} {\bibfnamefont {P.~O.}\ \bibnamefont
  {et~al.}},\ }\href {\doibase 10.5281/zenodo.3949207} {\  (\bibinfo {year}
  {2020}),\ 10.5281/zenodo.3949207}\BibitemShut {NoStop}%
\bibitem [{\citenamefont {James}\ and\ \citenamefont
  {Roos}(1975)}]{James:1975dr}%
  \BibitemOpen
  \bibfield  {author} {\bibinfo {author} {\bibfnamefont {F.}~\bibnamefont
  {James}}\ and\ \bibinfo {author} {\bibfnamefont {M.}~\bibnamefont {Roos}},\
  }\href {\doibase 10.1016/0010-4655(75)90039-9} {\bibfield  {journal}
  {\bibinfo  {journal} {Comput. Phys. Commun.}\ }\textbf {\bibinfo {volume}
  {10}},\ \bibinfo {pages} {343} (\bibinfo {year} {1975})}\BibitemShut
  {NoStop}%
\bibitem [{\citenamefont {He}\ \emph {et~al.}(2024)\citenamefont {He},
  \citenamefont {Ma}, \citenamefont {Schmidt}, \citenamefont {Valencia},\ and\
  \citenamefont {Volkas}}]{He:2024iju}%
  \BibitemOpen
  \bibfield  {author} {\bibinfo {author} {\bibfnamefont {X.-G.}\ \bibnamefont
  {He}}, \bibinfo {author} {\bibfnamefont {X.-D.}\ \bibnamefont {Ma}}, \bibinfo
  {author} {\bibfnamefont {M.~A.}\ \bibnamefont {Schmidt}}, \bibinfo {author}
  {\bibfnamefont {G.}~\bibnamefont {Valencia}}, \ and\ \bibinfo {author}
  {\bibfnamefont {R.~R.}\ \bibnamefont {Volkas}},\ }\href@noop {} {\  (\bibinfo
  {year} {2024})},\ \Eprint {http://arxiv.org/abs/2403.12485} {arXiv:2403.12485
  [hep-ph]} \BibitemShut {NoStop}%
\bibitem [{\citenamefont {Badin}\ and\ \citenamefont
  {Petrov}(2010)}]{Badin:2010uh}%
  \BibitemOpen
  \bibfield  {author} {\bibinfo {author} {\bibfnamefont {A.}~\bibnamefont
  {Badin}}\ and\ \bibinfo {author} {\bibfnamefont {A.~A.}\ \bibnamefont
  {Petrov}},\ }\href {\doibase 10.1103/PhysRevD.82.034005} {\bibfield
  {journal} {\bibinfo  {journal} {Phys. Rev. D}\ }\textbf {\bibinfo {volume}
  {82}},\ \bibinfo {pages} {034005} (\bibinfo {year} {2010})},\ \Eprint
  {http://arxiv.org/abs/1005.1277} {arXiv:1005.1277 [hep-ph]} \BibitemShut
  {NoStop}%
\bibitem [{\citenamefont {Bauer}\ \emph {et~al.}(2022)\citenamefont {Bauer},
  \citenamefont {Neubert}, \citenamefont {Renner}, \citenamefont {Schnubel},\
  and\ \citenamefont {Thamm}}]{Bauer:2021mvw}%
  \BibitemOpen
  \bibfield  {author} {\bibinfo {author} {\bibfnamefont {M.}~\bibnamefont
  {Bauer}}, \bibinfo {author} {\bibfnamefont {M.}~\bibnamefont {Neubert}},
  \bibinfo {author} {\bibfnamefont {S.}~\bibnamefont {Renner}}, \bibinfo
  {author} {\bibfnamefont {M.}~\bibnamefont {Schnubel}}, \ and\ \bibinfo
  {author} {\bibfnamefont {A.}~\bibnamefont {Thamm}},\ }\href {\doibase
  10.1007/JHEP09(2022)056} {\bibfield  {journal} {\bibinfo  {journal} {JHEP}\
  }\textbf {\bibinfo {volume} {09}},\ \bibinfo {pages} {056} (\bibinfo {year}
  {2022})},\ \Eprint {http://arxiv.org/abs/2110.10698} {arXiv:2110.10698
  [hep-ph]} \BibitemShut {NoStop}%
\bibitem [{\citenamefont {Li}\ and\ \citenamefont
  {Tandean}(2023)}]{Li:2023sjf}%
  \BibitemOpen
  \bibfield  {author} {\bibinfo {author} {\bibfnamefont {G.}~\bibnamefont
  {Li}}\ and\ \bibinfo {author} {\bibfnamefont {J.}~\bibnamefont {Tandean}},\
  }\href {\doibase 10.1007/JHEP11(2023)205} {\bibfield  {journal} {\bibinfo
  {journal} {JHEP}\ }\textbf {\bibinfo {volume} {11}},\ \bibinfo {pages} {205}
  (\bibinfo {year} {2023})},\ \Eprint {http://arxiv.org/abs/2306.05333}
  {arXiv:2306.05333 [hep-ph]} \BibitemShut {NoStop}%
\bibitem [{\citenamefont {Cortina~Gil}\ \emph {et~al.}(2021)\citenamefont
  {Cortina~Gil} \emph {et~al.}}]{NA62:2021zjw}%
  \BibitemOpen
  \bibfield  {author} {\bibinfo {author} {\bibfnamefont {E.}~\bibnamefont
  {Cortina~Gil}} \emph {et~al.} (\bibinfo {collaboration} {NA62}),\ }\href
  {\doibase 10.1007/JHEP06(2021)093} {\bibfield  {journal} {\bibinfo  {journal}
  {JHEP}\ }\textbf {\bibinfo {volume} {06}},\ \bibinfo {pages} {093} (\bibinfo
  {year} {2021})},\ \Eprint {http://arxiv.org/abs/2103.15389} {arXiv:2103.15389
  [hep-ex]} \BibitemShut {NoStop}%
\bibitem [{\citenamefont {Ahn}\ \emph {et~al.}(2021)\citenamefont {Ahn} \emph
  {et~al.}}]{KOTO:2020prk}%
  \BibitemOpen
  \bibfield  {author} {\bibinfo {author} {\bibfnamefont {J.~K.}\ \bibnamefont
  {Ahn}} \emph {et~al.} (\bibinfo {collaboration} {KOTO}),\ }\href {\doibase
  10.1103/PhysRevLett.126.121801} {\bibfield  {journal} {\bibinfo  {journal}
  {Phys. Rev. Lett.}\ }\textbf {\bibinfo {volume} {126}},\ \bibinfo {pages}
  {121801} (\bibinfo {year} {2021})},\ \Eprint
  {http://arxiv.org/abs/2012.07571} {arXiv:2012.07571 [hep-ex]} \BibitemShut
  {NoStop}%
\bibitem [{\citenamefont {Ablikim}\ \emph {et~al.}(2022)\citenamefont {Ablikim}
  \emph {et~al.}}]{BESIII:2021slf}%
  \BibitemOpen
  \bibfield  {author} {\bibinfo {author} {\bibfnamefont {M.}~\bibnamefont
  {Ablikim}} \emph {et~al.} (\bibinfo {collaboration} {BESIII}),\ }\href
  {\doibase 10.1103/PhysRevD.105.L071102} {\bibfield  {journal} {\bibinfo
  {journal} {Phys. Rev. D}\ }\textbf {\bibinfo {volume} {105}},\ \bibinfo
  {pages} {L071102} (\bibinfo {year} {2022})},\ \Eprint
  {http://arxiv.org/abs/2112.14236} {arXiv:2112.14236 [hep-ex]} \BibitemShut
  {NoStop}%
\bibitem [{\citenamefont {Khachatryan}\ \emph {et~al.}(2015)\citenamefont
  {Khachatryan} \emph {et~al.}}]{CMS:2014ofj}%
  \BibitemOpen
  \bibfield  {author} {\bibinfo {author} {\bibfnamefont {V.}~\bibnamefont
  {Khachatryan}} \emph {et~al.} (\bibinfo {collaboration} {CMS}),\ }\href
  {\doibase 10.1103/PhysRevLett.114.101801} {\bibfield  {journal} {\bibinfo
  {journal} {Phys. Rev. Lett.}\ }\textbf {\bibinfo {volume} {114}},\ \bibinfo
  {pages} {101801} (\bibinfo {year} {2015})},\ \Eprint
  {http://arxiv.org/abs/1410.1149} {arXiv:1410.1149 [hep-ex]} \BibitemShut
  {NoStop}%
\bibitem [{\citenamefont {Aad}\ \emph {et~al.}(2024)\citenamefont {Aad} \emph
  {et~al.}}]{ATLAS:2024vqf}%
  \BibitemOpen
  \bibfield  {author} {\bibinfo {author} {\bibfnamefont {G.}~\bibnamefont
  {Aad}} \emph {et~al.} (\bibinfo {collaboration} {ATLAS}),\ }\href@noop {} {\
  (\bibinfo {year} {2024})},\ \Eprint {http://arxiv.org/abs/2403.02793}
  {arXiv:2403.02793 [hep-ex]} \BibitemShut {NoStop}%
\bibitem [{\citenamefont {Bauer}\ \emph {et~al.}(2021)\citenamefont {Bauer},
  \citenamefont {Neubert}, \citenamefont {Renner}, \citenamefont {Schnubel},\
  and\ \citenamefont {Thamm}}]{Bauer:2020jbp}%
  \BibitemOpen
  \bibfield  {author} {\bibinfo {author} {\bibfnamefont {M.}~\bibnamefont
  {Bauer}}, \bibinfo {author} {\bibfnamefont {M.}~\bibnamefont {Neubert}},
  \bibinfo {author} {\bibfnamefont {S.}~\bibnamefont {Renner}}, \bibinfo
  {author} {\bibfnamefont {M.}~\bibnamefont {Schnubel}}, \ and\ \bibinfo
  {author} {\bibfnamefont {A.}~\bibnamefont {Thamm}},\ }\href {\doibase
  10.1007/JHEP04(2021)063} {\bibfield  {journal} {\bibinfo  {journal} {JHEP}\
  }\textbf {\bibinfo {volume} {04}},\ \bibinfo {pages} {063} (\bibinfo {year}
  {2021})},\ \Eprint {http://arxiv.org/abs/2012.12272} {arXiv:2012.12272
  [hep-ph]} \BibitemShut {NoStop}%
\bibitem [{\citenamefont {Buchalla}\ and\ \citenamefont
  {Buras}(1994)}]{Buchalla:1993wq}%
  \BibitemOpen
  \bibfield  {author} {\bibinfo {author} {\bibfnamefont {G.}~\bibnamefont
  {Buchalla}}\ and\ \bibinfo {author} {\bibfnamefont {A.~J.}\ \bibnamefont
  {Buras}},\ }\href {\doibase 10.1016/0550-3213(94)90496-0} {\bibfield
  {journal} {\bibinfo  {journal} {Nucl. Phys. B}\ }\textbf {\bibinfo {volume}
  {412}},\ \bibinfo {pages} {106} (\bibinfo {year} {1994})},\ \Eprint
  {http://arxiv.org/abs/hep-ph/9308272} {arXiv:hep-ph/9308272} \BibitemShut
  {NoStop}%
\bibitem [{\citenamefont {Misiak}\ and\ \citenamefont
  {Urban}(1999)}]{Misiak:1999yg}%
  \BibitemOpen
  \bibfield  {author} {\bibinfo {author} {\bibfnamefont {M.}~\bibnamefont
  {Misiak}}\ and\ \bibinfo {author} {\bibfnamefont {J.}~\bibnamefont {Urban}},\
  }\href {\doibase 10.1016/S0370-2693(99)00150-1} {\bibfield  {journal}
  {\bibinfo  {journal} {Phys. Lett. B}\ }\textbf {\bibinfo {volume} {451}},\
  \bibinfo {pages} {161} (\bibinfo {year} {1999})},\ \Eprint
  {http://arxiv.org/abs/hep-ph/9901278} {arXiv:hep-ph/9901278} \BibitemShut
  {NoStop}%
\bibitem [{\citenamefont {Buchalla}\ and\ \citenamefont
  {Buras}(1999)}]{Buchalla:1998ba}%
  \BibitemOpen
  \bibfield  {author} {\bibinfo {author} {\bibfnamefont {G.}~\bibnamefont
  {Buchalla}}\ and\ \bibinfo {author} {\bibfnamefont {A.~J.}\ \bibnamefont
  {Buras}},\ }\href {\doibase 10.1016/S0550-3213(99)00149-2} {\bibfield
  {journal} {\bibinfo  {journal} {Nucl. Phys. B}\ }\textbf {\bibinfo {volume}
  {548}},\ \bibinfo {pages} {309} (\bibinfo {year} {1999})},\ \Eprint
  {http://arxiv.org/abs/hep-ph/9901288} {arXiv:hep-ph/9901288} \BibitemShut
  {NoStop}%
\bibitem [{\citenamefont {Brod}\ \emph {et~al.}(2011)\citenamefont {Brod},
  \citenamefont {Gorbahn},\ and\ \citenamefont {Stamou}}]{Brod:2010hi}%
  \BibitemOpen
  \bibfield  {author} {\bibinfo {author} {\bibfnamefont {J.}~\bibnamefont
  {Brod}}, \bibinfo {author} {\bibfnamefont {M.}~\bibnamefont {Gorbahn}}, \
  and\ \bibinfo {author} {\bibfnamefont {E.}~\bibnamefont {Stamou}},\ }\href
  {\doibase 10.1103/PhysRevD.83.034030} {\bibfield  {journal} {\bibinfo
  {journal} {Phys. Rev. D}\ }\textbf {\bibinfo {volume} {83}},\ \bibinfo
  {pages} {034030} (\bibinfo {year} {2011})},\ \Eprint
  {http://arxiv.org/abs/1009.0947} {arXiv:1009.0947 [hep-ph]} \BibitemShut
  {NoStop}%
\bibitem [{\citenamefont {Bhattacharya}\ \emph {et~al.}(2019)\citenamefont
  {Bhattacharya}, \citenamefont {Grant},\ and\ \citenamefont
  {Petrov}}]{Bhattacharya:2018msv}%
  \BibitemOpen
  \bibfield  {author} {\bibinfo {author} {\bibfnamefont {B.}~\bibnamefont
  {Bhattacharya}}, \bibinfo {author} {\bibfnamefont {C.~M.}\ \bibnamefont
  {Grant}}, \ and\ \bibinfo {author} {\bibfnamefont {A.~A.}\ \bibnamefont
  {Petrov}},\ }\href {\doibase 10.1103/PhysRevD.99.093010} {\bibfield
  {journal} {\bibinfo  {journal} {Phys. Rev. D}\ }\textbf {\bibinfo {volume}
  {99}},\ \bibinfo {pages} {093010} (\bibinfo {year} {2019})},\ \Eprint
  {http://arxiv.org/abs/1809.04606} {arXiv:1809.04606 [hep-ph]} \BibitemShut
  {NoStop}%
\bibitem [{\citenamefont {Parrott}\ \emph {et~al.}(2023)\citenamefont
  {Parrott}, \citenamefont {Bouchard},\ and\ \citenamefont
  {Davies}}]{Parrott:2022rgu}%
  \BibitemOpen
  \bibfield  {author} {\bibinfo {author} {\bibfnamefont {W.~G.}\ \bibnamefont
  {Parrott}}, \bibinfo {author} {\bibfnamefont {C.}~\bibnamefont {Bouchard}}, \
  and\ \bibinfo {author} {\bibfnamefont {C.~T.~H.}\ \bibnamefont {Davies}}
  (\bibinfo {collaboration} {(HPQCD collaboration)\textsection{}, HPQCD}),\
  }\href {\doibase 10.1103/PhysRevD.107.014510} {\bibfield  {journal} {\bibinfo
   {journal} {Phys. Rev. D}\ }\textbf {\bibinfo {volume} {107}},\ \bibinfo
  {pages} {014510} (\bibinfo {year} {2023})},\ \Eprint
  {http://arxiv.org/abs/2207.12468} {arXiv:2207.12468 [hep-lat]} \BibitemShut
  {NoStop}%
\bibitem [{\citenamefont {Horgan}\ \emph {et~al.}(2014)\citenamefont {Horgan},
  \citenamefont {Liu}, \citenamefont {Meinel},\ and\ \citenamefont
  {Wingate}}]{Horgan:2013hoa}%
  \BibitemOpen
  \bibfield  {author} {\bibinfo {author} {\bibfnamefont {R.~R.}\ \bibnamefont
  {Horgan}}, \bibinfo {author} {\bibfnamefont {Z.}~\bibnamefont {Liu}},
  \bibinfo {author} {\bibfnamefont {S.}~\bibnamefont {Meinel}}, \ and\ \bibinfo
  {author} {\bibfnamefont {M.}~\bibnamefont {Wingate}},\ }\href {\doibase
  10.1103/PhysRevD.89.094501} {\bibfield  {journal} {\bibinfo  {journal} {Phys.
  Rev. D}\ }\textbf {\bibinfo {volume} {89}},\ \bibinfo {pages} {094501}
  (\bibinfo {year} {2014})},\ \Eprint {http://arxiv.org/abs/1310.3722}
  {arXiv:1310.3722 [hep-lat]} \BibitemShut {NoStop}%
\bibitem [{\citenamefont {Horgan}\ \emph {et~al.}(2015)\citenamefont {Horgan},
  \citenamefont {Liu}, \citenamefont {Meinel},\ and\ \citenamefont
  {Wingate}}]{Horgan:2015vla}%
  \BibitemOpen
  \bibfield  {author} {\bibinfo {author} {\bibfnamefont {R.~R.}\ \bibnamefont
  {Horgan}}, \bibinfo {author} {\bibfnamefont {Z.}~\bibnamefont {Liu}},
  \bibinfo {author} {\bibfnamefont {S.}~\bibnamefont {Meinel}}, \ and\ \bibinfo
  {author} {\bibfnamefont {M.}~\bibnamefont {Wingate}},\ }\href {\doibase
  10.22323/1.214.0372} {\bibfield  {journal} {\bibinfo  {journal} {PoS}\
  }\textbf {\bibinfo {volume} {LATTICE2014}},\ \bibinfo {pages} {372} (\bibinfo
  {year} {2015})},\ \Eprint {http://arxiv.org/abs/1501.00367} {arXiv:1501.00367
  [hep-lat]} \BibitemShut {NoStop}%
\bibitem [{\citenamefont {Gubernari}\ \emph {et~al.}(2019)\citenamefont
  {Gubernari}, \citenamefont {Kokulu},\ and\ \citenamefont {van
  Dyk}}]{Gubernari:2018wyi}%
  \BibitemOpen
  \bibfield  {author} {\bibinfo {author} {\bibfnamefont {N.}~\bibnamefont
  {Gubernari}}, \bibinfo {author} {\bibfnamefont {A.}~\bibnamefont {Kokulu}}, \
  and\ \bibinfo {author} {\bibfnamefont {D.}~\bibnamefont {van Dyk}},\ }\href
  {\doibase 10.1007/JHEP01(2019)150} {\bibfield  {journal} {\bibinfo  {journal}
  {JHEP}\ }\textbf {\bibinfo {volume} {01}},\ \bibinfo {pages} {150} (\bibinfo
  {year} {2019})},\ \Eprint {http://arxiv.org/abs/1811.00983} {arXiv:1811.00983
  [hep-ph]} \BibitemShut {NoStop}%
\bibitem [{\citenamefont {Bharucha}\ \emph {et~al.}(2016)\citenamefont
  {Bharucha}, \citenamefont {Straub},\ and\ \citenamefont
  {Zwicky}}]{Bharucha:2015bzk}%
  \BibitemOpen
  \bibfield  {author} {\bibinfo {author} {\bibfnamefont {A.}~\bibnamefont
  {Bharucha}}, \bibinfo {author} {\bibfnamefont {D.~M.}\ \bibnamefont
  {Straub}}, \ and\ \bibinfo {author} {\bibfnamefont {R.}~\bibnamefont
  {Zwicky}},\ }\href {\doibase 10.1007/JHEP08(2016)098} {\bibfield  {journal}
  {\bibinfo  {journal} {JHEP}\ }\textbf {\bibinfo {volume} {08}},\ \bibinfo
  {pages} {098} (\bibinfo {year} {2016})},\ \Eprint
  {http://arxiv.org/abs/1503.05534} {arXiv:1503.05534 [hep-ph]} \BibitemShut
  {NoStop}%
\bibitem [{\citenamefont {Aoki}\ \emph {et~al.}(2022)\citenamefont {Aoki} \emph
  {et~al.}}]{FlavourLatticeAveragingGroupFLAG:2021npn}%
  \BibitemOpen
  \bibfield  {author} {\bibinfo {author} {\bibfnamefont {Y.}~\bibnamefont
  {Aoki}} \emph {et~al.} (\bibinfo {collaboration} {Flavour Lattice Averaging
  Group (FLAG)}),\ }\href {\doibase 10.1140/epjc/s10052-022-10536-1} {\bibfield
   {journal} {\bibinfo  {journal} {Eur. Phys. J. C}\ }\textbf {\bibinfo
  {volume} {82}},\ \bibinfo {pages} {869} (\bibinfo {year} {2022})},\ \Eprint
  {http://arxiv.org/abs/2111.09849} {arXiv:2111.09849 [hep-lat]} \BibitemShut
  {NoStop}%
\bibitem [{\citenamefont {Adachi}\ \emph {et~al.}()\citenamefont {Adachi} \emph
  {et~al.}}]{Belle-II:PrivateCommunication}%
  \BibitemOpen
  \bibfield  {author} {\bibinfo {author} {\bibfnamefont {I.}~\bibnamefont
  {Adachi}} \emph {et~al.} (\bibinfo {collaboration} {Belle-II}),\ }\href@noop
  {} {}\bibinfo {howpublished} {{Private Communication}}\BibitemShut {NoStop}%
\end{thebibliography}%

\end{document}